\documentclass[preprint]{imsart}

\RequirePackage[OT1]{fontenc}
\RequirePackage{amsthm,amsmath}
\RequirePackage[numbers]{natbib}
\RequirePackage[colorlinks,citecolor=blue,urlcolor=blue]{hyperref}
\arxiv{arXiv:0000.0000}

\startlocaldefs
\numberwithin{equation}{section}
\numberwithin{table}{section}
\theoremstyle{plain}
\newtheorem{thm}{Theorem}[section]

\newtheorem{lem}{Lemma}[section]
\endlocaldefs

\usepackage{pdflscape}

\usepackage{comment}
\usepackage{eucal} 
\usepackage{latexsym} 
\usepackage{amssymb} 
\usepackage{amsmath}
\usepackage{accents}
\usepackage{enumitem}
\usepackage{subcaption}
\usepackage{cancel}
\usepackage{rotating}
\usepackage{url}
\usepackage{bigints}

\usepackage{lscape}

\usepackage{ae,lmodern}  
\usepackage[french, english]{babel}
\usepackage[utf8]{inputenc}
\usepackage[T1]{fontenc}
\usepackage{booktabs}

\usepackage{multirow}

\usepackage{float}

\newcommand{\SSbb}{\ensuremath{\mathbf{S}}}

\newcommand{\Xb}{\ensuremath{\mathbf{X}}}
  
\newcommand{\xb}{\ensuremath{\mathbf{x}}} 
\newcommand{\yb}{\ensuremath{\mathbf{y}}} 
 
\newcommand{\n}{^{(n)}}
\newcommand{\thetab}{\ensuremath{\pmb{\theta}}}
\newcommand{\betab}{\ensuremath{\pmb{\beta}}}
\newcommand{\gammab}{\ensuremath{\pmb{\gamma}}}
\newcommand{\deltab}{\ensuremath{\pmb{\delta}}}
\newcommand{\varthetab}{\ensuremath{\pmb{\vartheta}}}  
\newcommand{\taub}{\ensuremath{\pmb{\tau}}} 
\newcommand{\lambdab}{\ensuremath{\pmb{\lambda}}} 
\newcommand{\Sigmab}{\ensuremath{\pmb{\Sigma}}}
\newcommand{\mub}{\ensuremath{\pmb{\mu}}}
\newcommand{\Deltab}{\ensuremath{\pmb{\Delta}}}

\newcommand{\Sb}{\ensuremath{\mathbf{S}}}
\newcommand{\Ib}{\ensuremath{\mathbf{I}}}
\newcommand{\zerob}{\ensuremath{\mathbf{0}}}

\newcommand{\bb}{\ensuremath{\mathbf{b}}}
\newcommand{\Ab}{\ensuremath{\mathbf{A}}}
\newcommand{\Hb}{\ensuremath{\mathbf{H}}}
\newcommand{\Rd}{\ensuremath{\mathbb{R}^d}} 
\newcommand{\Ik}{\ensuremath{\mathbf{I}_d}} 
\newcommand{\Idd}{\ensuremath{\mathbf{I}_{d^2}}} 
\newcommand{\Jd}{\ensuremath{\mathbf{J}_d}} 
\newcommand{\Kd}{\ensuremath{\mathbf{K}_d}}

\newcommand{\Pd}{\ensuremath{\mathbf{P}_d}}
\newcommand{\Rzp}{\ensuremath{\mathbb{R}^{+}_0}} 
\newcommand{\Ub}{\ensuremath{\mathbf{U}}} 
 
\newcommand{\eb}{\ensuremath{\mathbf{e}}} 
\newcommand{\tb}{\ensuremath{\mathbf{t}}} 
 
\newcommand{\vect}{\ensuremath{\textrm{vec}}} 
\newcommand{\vech}{\ensuremath{\textrm{vech}}}

\newcommand{\Ubin}{\ensuremath{\mathbf{U}_i}}
\newcommand{\Ubjn}{\ensuremath{\mathbf{U}_j}}

\newcommand{\din}{\ensuremath{d_i}} 
\newcommand{\djn}{\ensuremath{d_j}} 

\newcommand{\Lambdab}{\ensuremath{\pmb{\Lambda}}} 
\newcommand{\Gammab}{\ensuremath{\pmb{\Gamma}}} 

\newcommand{\R}{\mathbb{R}}

\begin{document}
\selectlanguage{english}

\begin{frontmatter}
\title{Optimal tests for elliptical symmetry:\\ specified and unspecified location
 }
\runtitle{Optimal tests for elliptical symmetry}

\begin{aug}
\author{\fnms{Sla\dj ana} \snm{Babi\'c}\thanksref{t1}\ead[label=e1]{sladana.babic@ugent.be}},
\author{\fnms{Laetitia} \snm{Gelbgras}\!
\ead[label=e2]{laetitia.gelbgras@yahoo.fr},}\\ 
\author{\fnms{Marc} \snm{Hallin}\ead[label=e3]{mhallin@ulb.ac.be}},
\and
\author{\fnms{Christophe} \snm{Ley}
\ead[label=e4]{christophe.ley@ugent.be}
}

\thankstext{t1}{Sla\dj ana Babi\' c was supported by a grant (165880) as a PhD Fellow of the Research Foundation-Flanders (FWO).}
\runauthor{Babi\'c, Gelbgras, Hallin, and Ley}


\address{Sla\dj ana Babi\'c\\
Department of Applied Mathematics, \\ Computer Science, and Statistics\\ 
Ghent University\\  Krijgslaan 281, S9\\
B-9000 Gent, Belgium\\
and\\ Vlerick Business School\\ 
\printead{e1}\\
}

\address{Laetitia Gelbgras\\
D\' epartement de Math\' ematique\\
Universit\' e libre de Bruxelles\\
Campus de la Plaine  CP210 \\
B-1050 Bruxelles,  Belgium \\ 
\printead{e2}\\
}

\address{Marc Hallin\\
ECARES and\\  D\' epartement de Math\' ematique\\
Universit\' e libre de Bruxelles CP114/4\\
Ave. F.D. Roosevelt, 50\\ 
B-1050 Bruxelles,  Belgium \\ 
\printead{e3}\\
}

\address{Christophe Ley\\
Department of Applied Mathematics, \\ Computer Science, and Statistics\\ 
Ghent University\\  Krijgslaan 281, S9\\
B-9000 Gent, Belgium\\
\printead{e4}\\
}
\end{aug}

\begin{abstract} \ Although the assumption of elliptical symmetry is quite common in multivariate analysis and widespread in a number of applications, the problem of testing the null hypothesis of ellipticity  so far has not been addressed in a fully satisfactory way. Most of the literature in the area  indeed addresses the null hypothesis of elliptical symmetry with specified location and actually addresses location rather than non-elliptical alternatives.  In this paper, we are proposing new classes of testing procedures, both for specified and unspecified location.  The backbone of our construction is  Le Cam's asymptotic  theory of statistical experiments, and optimality is to be understood locally and asymptotically within the family of generalized skew-elliptical distributions. The tests we are proposing are meeting all the desired properties of a``good'' test of elliptical symmetry:  they have a simple asymptotic   distribution under the entire null hypothesis of elliptical symmetry with unspecified radial density and shape parameter;  they are affine-invariant, computationally fast, intuitively understandable, and not too demanding in terms of  moments. While achieving optimality against generalized skew-elliptical alternatives, they remain quite powerful under a much broader class of non-elliptical distributions and significantly outperform the available competitors.  
\end{abstract}



\end{frontmatter}

\section{Introduction}
\subsection{The ubiquitous assumption of elliptical symmetry}

Elliptical symmetry is a fundamental structural assumption in multivariate analysis and econometrics. It has been popularized in the 1970's as a natural extension of the (overly restrictive) multinormal assumption. Since then, most  multivariate analysis  procedures have been extended under elliptical symmetry with unspecified and sometimes possibly heavy-tailed {\it radial density} (see below for a definition): 
one- and $K$-sample location and shape problems   (\citep{UmRandles98, hallin2002optimal, hallin2006semiparametrically1, hallin2006semiparametrically2, HPd2008AoS, HPd2009JMVA}), 
serial dependence and  time series (\citep{HPd2002b, HPd2004AoS, HPd2004b}), 
linear models with VARMA errors (\citep{HPd2005JMVA}), 
one- and $K$-sample  principal component problems   (\citep{hallin2010optimal, HPdV2010JNPS, HPdV2013Bernou, HPdV2014JASA}),  to cite but a few.  Most tests proposed in those references are either pseudo-Gaussian or based on variations of {\it Mahalanobis} ranks and signs, {\it interdirections}, etc.   
Elliptical densities also are considered in capital asset pricing models \citep{hodgson2002testing}, semiparametric density estimation \citep{liebscher2005semiparametric}, graphical models \citep{vogel2011elliptical},   multivariate tail estimation \citep{dominicy2017multivariate}, and many other areas. 


Let $\Xb_1,\ldots,\Xb_n$ denote a sample of $n$ i.i.d.~$d$-dimensional observations. A $d$-dimensional random vector $\Xb$ is said to be elliptically symmetric about some location parameter $\thetab\in\R^d$ if its density~$\underline{f}$ is of the form 
\begin{equation}\label{def1}
\xb\mapsto{\underline{f}(\xb;\thetab,\Sigmab,f)}=c_{d,f}|\Sigmab|^{-1/2}f\left(\|\Sigmab^{-1/2}(\xb-\thetab)\|\right),\qquad\mathbf{x}\in\mathbb{R}^d,
\end{equation}
where $\Sigmab\in\mathcal{S}_d$ (the class of symmetric positive definite real $d\times d$ matrices) is a {\it scatter} parameter, $f:\mathbb{R}^+_0\to\mathbb{R}^+$ is an a.e.\  strictly positive function called {\it radial density}, and $c_{d,f}$ is a normalizing constant depending on~$f$ and the dimension~$d$. Well-known instances 
 are the multivariate normal, Student $t$ and power-exponential distributions. The family of elliptical distributions has several appealing properties. For instance, it is closed under affine transformations, and its marginal and conditional distributions are also elliptically symmetric: see \citep{paindaveine2014elliptical} for details. A   salient feature is the  stochastic representation of elliptical variables: an elliptically symmetric random vector $\Xb$ is conveniently   represented as
\begin{equation}\label{ellstoch}
\Xb =_{d} \thetab + \rho\Lambdab \Ub^{(r)},
\end{equation}
where $=_{d}$ stands for equality in distribution,  $\Lambdab \in \mathbb{R}^{d\times r}$ has   rank $r\leq d$ and is such that $\Lambdab \Lambdab' = \Sigmab$,  $\Ub^{(r)}$ is an $r$-dimensional random vector uniformly distributed over the unit hypersphere, and $\rho$ is a nonnegative random variable independent of $ \Ub^{(r)}.$ Letting $\mu_{\ell,f}:=\int_0^\infty r^\ell f(r)dr$, the density of~$\rho$ is
\begin{equation}\label{ftilde}
r\mapsto\tilde{f}_d(r):=  \mu_{d-1,f}^{-1} r^{d-1}f(r), \qquad r>0.
\end{equation}
 The existence of this density thus requires $\mu_{d-1,f}$ to be finite, and $\Xb$ admits finite moments of order $\alpha>0$ if and only if $\mu_{d+\alpha-1,f}<\infty$. Inference in elliptically symmetric distributions has been abundantly studied:  see \citep{paindaveine2014elliptical} for a survey.

\subsection{Testing for elliptical symmetry}

Considering the omnipresence of the  assumption of elliptical symmetry, it is of primary importance to be able to test whether  that assumption actually holds true, and various   tests have been proposed in the literature for this problem. We  briefly mention the most popular of them, along with their respective  pitfalls; later on,  we will focus more closely on those   used in our comparative  Monte Carlo study (Section~\ref{numsec}). We also mention tests for spherical symmetry,  a special case of elliptical symmetry corresponding to $\mub=\zerob$ and $\Sigmab={\bf I}_d$, the $d\times d$ identity matrix. These tests  in principle can be turned into elliptical symmetry tests by standardizing the data via $\hat\Sigmab^{-1/2}(\Xb_i-\hat\thetab)$ where $\hat\thetab$ and $\hat\Sigmab$ are location and scatter estimators. 
\begin{itemize}
\item[(i)] \citet{beran1979testing} introduces a test based on marginal signs and ranks. That test is neither distribution-free nor   affine-invariant; moreover,  there are no practical guidelines   to the choice of the basis functions involved in the test statistic.
\item[(ii)] \citet{baringhaus1991testing} proposes a Cram\'er-von Mises type test for spherical symmetry based on the independence between norm and direction. It assumes the location parameter to be known and its asymptotic distribution is not simple to use.  Dyckerhoff et al.\  \cite{dyckerhoff2015depth} have shown by simulations that this test can be used as a test for elliptical symmetry in dimension 2.
\item[(iii)] \citet{koltchinskii2000testing} consider bootstrap-type tests based on a class of functions closed under orthogonal transformations. Their tests have no known asymptotic distribution, which is why a bootstrap procedure is required to get the critical values. 
\item[(iv)] Manzotti et al.~\citep{manzotti2002statistic} develop a test based on spherical harmonics to test whether the  standardized
 vectors $\hat\Sigmab^{-1/2}(\Xb_i-\hat\thetab)/||\hat\Sigmab^{-1/2}(\Xb_i-\hat\thetab)||$ are uniformly distributed on the unit sphere. The test is computationally demanding and requires moments of order $4$. 
\item[(v)] \citet{schott2002testing} builds a Wald-type test to compare the sample fourth-order moments with the expected theoretical ones under elliptical symmetry. Being based on fourth-order moments, the test is very simple to use but requires moments of order $8$. Moreover, it has very low power against several alternatives. 
\item[(vi)] \citet{huffer2007test} propose a Pearson chi-square type test with multi-dimensional cells. Its asymptotic distribution exists only in case of normality, otherwise bootstrap techniques are required. 
\item[(vii)] \citet{Cassart07} and Cassart et al.\,\citep{CHPd2008JSPI} construct a pseudo-Gaussian test that is most efficient against  a multivariate form of Fechner-type  asymmetry. The test requires finite moments of order $4$.  
\end{itemize}
Tests based on Monte Carlo simulations can be found in \citet{diks1999test} and \citet{zhu2000nonparametric};  \citet{li1997some} recur to graphical methods and \citet{zhu2004} build conditional tests. We refer  the reader to  \citet{serfling2004multivariate} and \citet{sakhanenko2008testing} for extensive reviews and  performance comparisons.

\subsection{Goal and organization of the   paper}

Despite the practical importance of the problem and the many proposals made in the literature, all tests for elliptical symmetry are suffering from some serious drawbacks. None of them, except for \citet{Cassart07}, is based on efficiency arguments; and, to the best of our  knowledge, none of them has been implemented in $\mathtt{R}$.  

 This paper is filling this gap by building  tests for elliptical symmetry that are optimal against the very popular class of \emph{generalized skew-elliptical distributions} which we  define more precisely in Section~\ref{sec:skewell}. It should be clear, however, that we never require the actual density of the observations to belong to that class, the choice of which is made because it encompasses many proposed skew distributions from the literature (see, e.g., \citet{genton2004skew1}). The $\mathtt{R}$-code is  available on request and an $\mathtt{R}$-package under preparation.

 The tests we are proposing are meeting all the desired properties of a ``good'' test of elliptical symmetry:  they have  simple asymptotic   distributions under the entire null hypothesis of elliptical symmetry with unspecified radial density and shape parameter;  they are affine-invariant, computationally fast, intuitively understandable, and not too demanding in terms of  moments. The latter property is particularly important when dealing with possibly heavy-tailed data as is often the case in a financial context. All our tests are devised for specified and, most importantly,  unspecified location parameter.  The latter indeed is the ``genuine''  problem here, as specified-location tests for ellipticity   typically run into major problems---see Section~\ref{pitsec} and the empirical illustration in Section~\ref{empirsec}.

The approach we are adopting   thus combines optimality and robustness concerns (distribution-freeness with respect to radial densities and minimal moment assumptions). The backbone of our construction is  Le Cam's asymptotic  theory of statistical experiments, and optimality is to be understood in the local asymptotic sense  (against local generalized skew-elliptical deviations from ellipticity). Under each scenario (specified and unspecified location), we  first build optimal parametric tests by assuming a given elliptical distribution. Then we make these tests  valid under the entire semiparametric family of elliptically symmetric distributions, while preserving their (parametric) optimality. 
As we shall see, 
 under specified location, the optimal  parametric test statistics do not involve the radial density, hence have all the same expression which consequently is {\it uniformly} optimal across radial densities---a rather rare phenomenon, which does not hold in other problems involving elliptical densities. When the location is unspecified, this uniform optimality property   gets lost, but we still obtain very simple and fast-to-compute test statistics that significantly outperform their competitors and do not require estimating the actual density, as is often the case.  A detailed  comparative study of the finite-sample performances of our tests is conducted in Section~\ref{numsec} and demonstrates the power of our procedures. 
  


The rest of the paper is organized as follows. In Section~\ref{mainsec:ULAN}, we describe the family of generalized skew-elliptical distributions and   state some mild conditions on the radial density $f$ which are required in order to establish  uniform local   asymptotic normality (ULAN) under given $f$. 
 In Section~\ref{sec:fix}, we derive, for given $f$,  the locally and asymptotically
optimal tests for  symmetry about a specified location $\thetab$. These tests are parametric, and valid under the known radial density $f$ only. We turn them into  semiparametric tests that remain valid under a broad class of radial  densities and, as already mentioned, also are uniformly optimal against alternatives involving the same class of densities.  Section~\ref{sec:notfix} deals with the unspecified location case,  for which  again we derive parametrically locally and asymptotically optimal tests,   which we turn   into  semiparametric ones, the properties of which we provide under the null and  contiguous alternatives. Asymptotic relative efficiencies with respect to the aforementioned pseudo-Gaussian test of \citet{Cassart07} are  calculated in Section~\ref{ARE}.
In Section~\ref{numsec}, we conduct a Monte Carlo simulation study of the finite-sample performances of the proposed tests and   their main competitors. Section~\ref{pitsec} stresses the all too often overlooked pitfalls of specified-location methods.  A real-data analysis is carried out in Section~\ref{empirsec} and conclusions are provided in Section~\ref{conclu}. Technical proofs are concentrated in the Appendix.

\section{Generalized skew-elliptical families and Uniform Local Asymptotic Normality (ULAN)}\label{mainsec:ULAN}

\subsection{Generalized skew-elliptical distributions}\label{sec:skewell}

As mentioned in the Introduction, our goal is to propose efficient tests  against a family of densities representative  of a broad class of skewed densities. The   family of {\it generalized skew-elliptical distributions} (\citep{genton2005generalized})  is an ideal candidate for this role.  

Let us assume that   the radial density $f$  in~\eqref{def1} belongs to 
$$
\mathcal{F}:=\left\{f:\mathbb{R}^+_0\to\mathbb{R}^+\,:\, f(r)>0 \,\mbox{ a.e.~and }\mu_{d-1;f}:=\int_0^\infty r^{d-1}f(r)dr<\infty\right\}.
$$
It is clear  from~\eqref{ellstoch} that $\rho$ and $\Sigmab$ are not separately identifiable, and we therefore impose a further identification constraint:  
\begin{equation}\label{idcond}
f\in\mathcal{F}_1:=\left\{f\in\mathcal{F}: \mu_{d-1;f}^{-1}\int_{\R^+}r^{d+1}f(r)dr=d\right\};
\end{equation}
Under this constraint, $\rho$ has  finite variance and~${\rm Cov}[\Xb]=\Sigmab$, which fully identifies  the scatter matrix $\Sigmab$. While  imposing the existence of finite second-order moments, \eqref{idcond}   does not imply any loss of generality, as second-order moments are needed anyway  (see Section~\ref{sec:ULAN})  to have   finite Fisher information  for skewness. It will be required in all statements involving ULAN (optimality, local powers, etc.), but is not necessary for   statements  made under the null hypothesis of ellipticity (mainly, the asymptotic size of a test and its  validity).  Gaussian densities clearly satisfy \eqref{idcond}, but  the Student ones do not,  and need to be rescaled. 

The {\it generalized skew-elliptical} alternatives we are interested in belong to the class of \emph{Azzalini-type distributions}. That class contains all generalizations  of the famous scalar skew-normal distribution introduced by \citet{azzalini1985class} with   density function $x\mapsto 2 \phi(x)\Phi(\lambda x), x\in\R,$ where $\phi$ and $\Phi$ stand for the standard normal density and distribution functions,  respectively, and~$\lambda\in\R$ is a skewness parameter. The idea underpinning the definition of the skew-normal consists in perturbating or modulating a \emph{symmetric kernel}, here the normal, by  multiplying it with a \emph{skewing function}, here $\Phi(\lambda x)$.  Its multivariate generalization was introduced in Azzalini and Dalla Valle~\citep{azzalini1996multivariate} by replacing the scalar normal density with the $d$-variate normal. \citet{azzalini1999statistical} and \citet{branco2001general} in turn extended the multivariate skew-normal into skew-elliptical distributions based on elliptically symmetric kernels. \citet{azzalini2003distributions} established a link between the distinct constructions of skew-elliptical distributions, extending them into a broader class of skewed distributions  very similar to the  generalized skew-elliptical distributions defined by \citet{genton2005generalized},
  with pdfs of the form 
\begin{align}\label{skewell}
&\xb\mapsto \underline{f}(\xb;\thetab,\Sigmab,\lambdab,f)  \\ 
&\qquad :=2\,c_{d,f}|\Sigmab|^{-1/2}f(\|\Sigmab^{-1/2}(\xb-\thetab)\|)\Pi(\lambdab'\Sigmab^{-1/2}(\xb-\thetab)), \ \xb\in\Rd,
\nonumber \end{align}
where $\thetab$, $\Sigmab$, $c_{d,f}$, and $f$ are defined as in \eqref{def1}; the  {\it skewing function}~$\Pi$ has values in $[0,1]$ and satisfies~$\Pi(-r)=1-\Pi(r)$ for   $r\in\R$; $\lambdab\in\Rd$\linebreak  plays the role of a \emph{skewness} parameter.
  The density~\eqref{skewell} thus results from perturbing the elliptically symmetric {\it kernel} $\underline{f} (\xb;\thetab,\Sigmab,f)$
   into~$2\underline{f}(\xb;\thetab,\Sigmab,f)\Pi(\lambdab'\Sigmab^{-1/2}(\xb-\thetab))$ 
  by multiplying it with a general skewing function $\Pi(\cdot)$; clearly,  the original symmetric version is retrieved for  $\lambdab=\zerob$. Typical choices for $\Pi$ are univariate distribution functions with symmetric densities, such as the normal or Student  ones; see the monograph by Genton~\citep{genton2004skew1}. We opted for this class of skew alternatives because of its popularity and  its ability to closely approximate a large variety of skewed distributions.

\subsection{Notation and some definitions}

Let $\Xb_1,\ldots,\Xb_n$ be i.i.d.~with   density~\eqref{skewell}. Denote by   ${\rm P}^{(n)}_{\thetab,\,\Sigmab,\,\lambdab;f,\Pi}$  the joint distribution of $(\Xb_1,\ldots,\Xb_n)$ which, in case $\lambdab=\zerob$, we simply write as~${\rm P}^{(n)}_{\thetab,\,\Sigmab,\,\zerob;f}$. Any couple $(f,\Pi)$ then induces a parametric location-scatter-skewness model
$$
\mathcal{\rm P}^{(n)}_{f,\Pi}:=\left\{{\rm P}^{(n)}_{\thetab,\,\Sigmab,\,\lambdab;f,\Pi}:\thetab\in\R^d,\Sigmab\in\mathcal{S}_d,\lambdab\in\Rd\right\}.
$$

We are   interested in testing $\mathcal{H}_0: \lambdab=\pmb{0}$ against~$\mathcal{H}_1: \lambdab\neq\pmb{0}$ in \eqref{skewell}, 
in the presence of a variety of 
 unspecified nuisances: $\Pi$ and/or $\thetab$ and/or~$\Sigmab$ and/or $f$ ... Depending on the case, the problem is either   
  parametric 
 or semiparametric.
The four types of testing problems we are considering   are
\begin{enumerate}
\item[(a)] (specified   $f$ and specified   $\thetab$)\vspace{-1.4mm}
\end{enumerate}
 $\mathcal{H}^{(n)}_{0;f,\thetab}:=\bigcup_{\,\Sigmab\in\mathcal{S}_d} {\rm P}^{(n)}_{\thetab,\,\Sigmab,\,\zerob;f}$ versus $\mathcal{H}^{(n)}_{1;f,\Pi,\thetab}:=\bigcup_{\,\Sigmab\in\mathcal{S}_d,\,\lambdab\neq \zerob} {\rm P}^{(n)}_{\thetab,\,\Sigmab,\,\lambdab;f\,,\Pi}$,
\begin{enumerate}
\item[(b)] (specified   $f$ and unspecified $\thetab$)\vspace{-2mm}
\end{enumerate}
 $\mathcal{H}^{(n)}_{0;f}:=\bigcup_{\,\thetab\in\Rd\!,\,\Sigmab\in\mathcal{S}_d} {\rm P}^{(n)}_{\thetab,\,\Sigmab,\,\zerob;f}$ versus $\mathcal{H}^{(n)}_{1;f,\Pi}=\bigcup_{\,\thetab\in\Rd\!,\,\Sigmab\in\mathcal{S}_d,\,\lambdab\neq \zerob} {\rm P}^{(n)}_{\thetab,\,\Sigmab,\,\lambdab;f,\Pi}$,
\begin{enumerate}
\item[(c)] (unspecified $f$  and specified $\thetab$)\vspace{-2mm}
\end{enumerate}
 $\mathcal{H}^{(n)}_{0;\thetab}:=\bigcup_{\,f\in\mathcal{F}_1,\,\Sigmab\in\mathcal{S}_d} {\rm P}^{(n)}_{\thetab,\,\Sigmab,\,\zerob;f}$ versus $\mathcal{H}^{(n)}_{1;\Pi,\thetab}:=\bigcup_{\,f\in\mathcal{F}_1,\,\Sigmab\in\mathcal{S}_d,\,\lambdab\neq \zerob} {\rm P}^{(n)}_{\thetab,\,\Sigmab,\,\lambdab;f,\Pi}$,  and
\begin{enumerate}
\item[(d)] (unspecified $f$ and unspecified $\thetab$)\vspace{-2mm}
\end{enumerate}
  $\mathcal{H}_{0}^{(n)}\!\!:=\!\bigcup_{\,f\in\mathcal{F}_1,\,\thetab\in\Rd\!,\,
\Sigmab\in\mathcal{S}_d}\! {\rm P}^{(n)}_{\thetab,\,\Sigmab,\,\zerob;f}\!\!$ versus $\mathcal{H}^{(n)}_{1;\Pi}\!:=\!\bigcup_{\,f\in\mathcal{F}_1,\,\thetab\in\Rd\!,\,\Sigmab\in\mathcal{S}_d,\lambdab\neq \zerob}\! {\rm P}^{(n)}_{\thetab,\,\Sigmab,\,\lambdab;f,\Pi}$;\smallskip

\noindent the skewing function $\Pi$ and the scatter $\Sigmab$ throughout remain unspecified.

For all $i=1,\ldots,n$,  denote by $\din(\thetab,\Sigmab):=\|\Sigmab^{-1/2}(\Xb_i-\thetab)\|$ the Mahalanobis  distance of $\Xb_i$ to $\thetab$ 
 and by 
 $\Ubin(\thetab,\Sigmab):=\Sigmab^{-1/2}(\Xb_i-\thetab)/\din(\thetab,\Sigmab)$ 
 its {\it multivariate sign}  in the metric $\Sigmab$. Under elliptical symmetry, those signs are uniformly distributed on the unit hypersphere of $\Rd$ whereas the radial quantities $\din(\thetab,\Sigmab)$ have common density $
\tilde{f}_d$, see \eqref{ftilde}. Any square root  of $\Sigmab$ 
 can be used in the previous definitions, but   we throughout denote by $\Sigmab^{1/2}$  the unique symmetric positive definite one.

Let $\Sb$ be a $d\times d$ symmetric matrix. We   throughout use the classical $\vect\Sb$ notation for the $d^2$-vector obtained by stacking the columns of $\Sb$ on top of each other and  write $\vech\mathbf{S}$ for the $d(d+1)/2$-dimensional vector stacking its   upper-triangular elements. We then denote by $\Pd$ the $(d(d+1)/2)\times d^2$ matrix such that $\Pd'(\vech\mathbf{S})=\vect\mathbf{S}$. Write  $\Sb^{\otimes 2}$ for the   Kronecker product~$\Sb\otimes\Sb$. Finally, denoting by~$\mathbf{e}_i$  the~$i^{th}$ vector of the canonical basis of $\Rd$,   define the~$d^2\times d^2$ {\it commutation matrix}    $\Kd:=\sum_{i,j=1}^d(\mathbf{e}_i\mathbf{e}_j')\otimes(\mathbf{e}_j\mathbf{e}_i')$
and the~$d^2\times d^2$ {\it projection matrix} $\Jd:=\sum_{i,j=1}^d(\mathbf{e}_i\mathbf{e}_j')\otimes(\mathbf{e}_i\mathbf{e}_j')=(\textrm{vec}\Ik)(\textrm{vec}\Ik)'.$ 

\subsection{Uniform Local Asymptotic Normality (ULAN)}\label{sec:ULAN}

The backbone of our construction of efficient tests   in the subsequent sections is the ULAN property, at $\lambdab=\zerob$, of the parametric model $\mathcal{\rm P}^{(n)}_{f,\Pi}$. This ULAN property  requires some further regularity conditions on  $f$. 
 Let $(\Omega,\mathcal{B}^d_{\Omega},\lambda)$ be a measure space, where $\lambda$ is a measure on the open subset $\Omega\subseteq\mathbb{R}^d$ equipped with its Borel $\sigma$-field $\mathcal{B}^d_{\Omega}$.  Denote by $L^2(\Omega,\lambda)$ the space of measurable func-\linebreak tions~$h:\Omega\to\mathbb{R}$ such that $\int_{\Omega}[h(\xb)]^2d\lambda(\xb)<\infty$,   by $L^2(\Rzp,\mu_j)$ the space of square-integrable functions with respect to the Lebesgue measure with weight $r^j$ over $\R_0^+$,  and by $L^2(\R,\nu_j)$ the space of square-integrable functions with respect to the Lebesgue measure with weight $e^{rj}$ over $\R$. We say that~$g\in L^2(\Omega,\lambda)$ admits a \emph{weak partial derivative} $T_i$ with respect to the $i^{th}$ variable iff
\begin{displaymath}
\int_{\Omega}g(\xb)\partial_i\varphi(\xb)d\xb=-\int_{\Omega}T_i(\xb)\varphi(\xb)d\xb
\end{displaymath}
for any function $\varphi\in C_0^{\infty}(\Omega)$, i.e. for any infinitely differentiable (in the classical sense) compactly supported function $\varphi$ on $\Omega$. If $T_i$ exists for all~$i$, the gradient $\mathbf{T}:=(T_1,\ldots,T_d)$ is also called the \emph{derivative of g in the sense of distributions} in $L^2(\Omega,\lambda)$. If, in addition, $\mathbf{T}\in L^2(\Omega,\lambda)$, then $g$ belongs to~$W^{1,2}(\Omega,\lambda)$, the \emph{Sobolev space of order 1} on $L^2(\Omega,\lambda)$. This space is a Banach space when  equipped with the norm $$\|g\|_{ W^{1,2}(\Omega,\lambda)}:=(\|g\|_{L^2(\Omega,\lambda)}^2+\sum_{i=1}^d\|T_i\|^2_{L^2(\Omega,\lambda)})^{1/2}.$$
 In particular, we will denote by $L^2(\Omega)$ and $W^{1,2}(\Omega)$ the case where $\lambda$ is the Lebesgue measure on $\Omega$.

With this in hand, let us state the  regularity assumptions we
 need for~ULAN.

{\sc Assumption (A1)}
The mapping $r\mapsto f^{1/2}(r)$ belongs to $W^{1,2}(\mathbb{R}_0,\mu_{d-1})$.

\noindent Define $\varphi_f(r):=-2(f^{1/2})'(r)/f^{1/2}(r)$, where $(f^{1/2})'$ stands for the weak derivative of $f^{1/2}$ in~$L^2(\mathbb{R}_0,\mu_{d-1})$. Assumption (A1) ensures finiteness of the {\it Fisher information for location} 
$$
\mathcal{I}_{d,f}:=c_{d,f}\int_{\Rd}\varphi_f^2(\|\xb\|)f(\|\xb\|)d\xb.
$$

{\sc Assumption (A2)}
The mapping $r\mapsto f^{1/2}_{\textrm{exp}}(r):=f^{1/2}(e^r)$ belongs to~$W^{1,2}(\mathbb{R}_0,\nu_{d})$.

\noindent Letting $\psi_f(r):=-2r^{-1}\left(f_{\rm exp}^{1/2}\right)^\prime(\log r)/f^{1/2}(r)$, where $(f_{\rm exp}^{1/2})'$ stands for the weak derivative of $f_{\rm exp}^{1/2}$ in $L^2(\mathbb{R}_0,\nu_{d})$, Assumption (A2) ensures finiteness of the {\it Fisher information for scatter}
$$
\mathcal{J}_{d,f}:=c_{d,f}\int_{\Rd}\|\xb\|^2\psi_f^2(\|\xb\|)f(\|\xb\|)d\xb.
$$
Now, if we assume the radial density $f$ to be continuously differentiable, then~$\varphi_f$ and $\psi_f$ both coincide with $-\dot{f}/f$ where  $\dot{f}$ is  the classical (strong) derivative of $f$.

%

Note that \eqref{idcond} is sufficient 
 for the finiteness of the {\it Fisher information for skewness} (see Theorem~\ref{ULAN} below), which only requires finite moments of order 2. 

 Finally, let  $\varthetab:=(\thetab',(\vech\Sigmab)',\lambdab')'$ and $\varthetab_{\pmb{0}}:=(\thetab',(\vech\Sigmab)',\pmb{0}')'$. We are now ready to state the ULAN property of the family $\mathcal{\rm P}^{(n)}_{f,\Pi}$ \emph{in the vicinity of symmetry}. 

\begin{thm}\label{ULAN}
Let $f\in\mathcal{F}_1$. Suppose that Assumptions (A1) and~(A2) hold, and that the skewing function $\Pi$ is continuously differentiable  at $0$, with derivative $\dot{\Pi}(0)\neq0$. Then, the family $\mathcal{\rm P}^{(n)}_{f,\Pi}$ is ULAN     at $\varthetab_0$ with respect to $\thetab$, $\Sigmab$ and $\lambdab$, with central sequence
\begin{align*}
&\Deltab_f(\varthetab_{\pmb{0}})=\begin{pmatrix} \Deltab_{f;1}(\varthetab_{\pmb{0}}) \\[2mm] \Deltab_{f;2}(\varthetab_{\pmb{0}}) \\[2mm] \Deltab_{3}(\varthetab_{\pmb{0}})
\end{pmatrix}\\
&:=\begin{pmatrix} \displaystyle n^{-1/2}\sum_{i=1}^n\varphi_f(\din(\thetab,\Sigmab))\Sigmab^{-1/2}\Ubin(\thetab,\Sigmab) \\ \displaystyle\frac{1}{2}n^{-1/2} \Pd(\Sigmab^{\otimes2})^{-1/2} \sum_{i=1}^n \emph{\vect}\left(\psi_f(\din(\thetab,\Sigmab))\din(\thetab,\Sigmab)\Ubin(\thetab,\Sigmab)\mathbf{U}_i^{'}(\thetab,\Sigmab)-\Ik \right) \\
\displaystyle2n^{-1/2}\dot{\Pi}(0)\sum_{i=1}^n\din(\thetab,\Sigmab)\Ubin(\thetab,\Sigmab)
\end{pmatrix}
\end{align*}
and Fisher information matrix
\begin{equation}\label{Fishinfo}
\Gammab_f(\varthetab_{\pmb{0}}):=\begin{pmatrix}
\Gammab_{f;11}(\varthetab_{\pmb{0}}) & \pmb{0} & \Gammab_{f;13}(\varthetab_{\pmb{0}}) \\ \pmb{0} & \Gammab_{f;22}(\varthetab_{\pmb{0}}) & \pmb{0} \\ \Gammab_{f;13}(\varthetab_{\pmb{0}}) & \pmb{0} & \Gammab_{f;33}(\varthetab_{\pmb{0}}) \end{pmatrix},
\end{equation}
where
$\displaystyle{\Gammab_{f;11}(\varthetab_{\pmb{0}}):=\frac{1}{d}\mathcal{I}_{d,f}\Sigmab^{-1}\!, \hspace{2mm}
\Gammab_{f;13}(\varthetab_{\pmb{0}}):=2\dot{\Pi}(0)\Sigmab^{-1/2}\!, \hspace{2mm}\Gammab_{f;33}(\varthetab_{\pmb{0}}):=4(\dot{\Pi}(0))^2\Ik,}$ 
and 
$\displaystyle{
\Gammab_{f;22}(\varthetab_{\pmb{0}}):=\frac{1}{4}\Pd(\Sigmab^{\otimes2})^{-1/2}\left[\frac{\mathcal{J}_{d,f}}{d(d+2)}(\Idd+\Kd+\Jd)-\Jd\right](\Sigmab^{\otimes2})^{-1/2}\Pd'.
}$

More precisely, for any sequence $\varthetab_{\pmb{0},n}=(\thetab_n^{'},(\emph{\vech}\Sigmab_n)',\pmb{0}')'$, where $\thetab_n~\!-~\!\thetab$ and~$\Sigmab_n-~\Sigmab$ are $O(n^{-1/2})$, and for any bounded sequence $\taub^{(n)}$ of the form~$((\mathbf{t}^{(n)})',(\emph{\vech}\Hb^{(n)})',({\boldsymbol\ell}^{(n)})')'=((\taub_1^{(n)})',(\taub_2^{(n)})',(\taub_3^{(n)})')'\in\mathbb{R}^{2d+d(d+1)/2}$, 
\begin{eqnarray*}
L^{(n)}_{\varthetab_{\pmb{0},n}+n^{-1/2}\taub^{(n)}/\varthetab_{\pmb{0},n};f}&:=&\log\left(\frac{d{\rm P}^{(n)}_{\varthetab_{\pmb{0},n}+n^{-1/2}\taub^{(n)};f,\Pi}}{d{\rm P}^{(n)}_{\varthetab_{\pmb{0},n};f}}\right)\\
&=&(\taub^{(n)})'\Deltab_f(\varthetab_{\pmb{0},n})-\frac{1}{2}(\taub^{(n)})'\Gammab_f(\varthetab_{\pmb{0}})\taub^{(n)}+o_{\rm P}(1)
\end{eqnarray*}
and
\begin{displaymath}
\Deltab_f(\varthetab_{\pmb{0},n})\stackrel{D}{\longrightarrow}\mathcal{N}_{2d+d(d+1)/2}(\pmb{0},\Gammab_f(\varthetab_{\pmb{0}}))
\end{displaymath}
under ${\rm P}^{(n)}_{\varthetab_{\pmb{0},n};f}$ as $n\to\infty$.
\label{ULAN}
\end{thm}

See Appendix~A for the proof.\smallskip   

Note that the central sequence for skewness $ \Deltab_{3}(\varthetab_{\pmb{0}})$ does not depend on~$f$; this, as we shall see,  has  strong implications on optimality properties.

An immediate consequence of the ULAN property is the  \emph{asymptotic li\-nearity}, as   $n \to \infty$, of the central sequence $\Deltab_f$ under ${\rm P}_{\varthetab_{\zerob};f}^{(n)}$:
\begin{equation}\label{aslin}
\Deltab_f(\varthetab_{\zerob}+n^{-1/2}\taub^{(n)}) - \Deltab_f(\varthetab_{\zerob}) = -\Gammab_f(\varthetab_{\pmb{0}})\taub^{(n)} + o_{\rm P}(1).
\end{equation}
 This 
 property classically plays a key role in the handling of nuisance parameters. Denote by $\hat\thetab\n$ and $\hat\Sigmab\n$  sequences of estimators of $\thetab$ and $\Sigmab$, respectively,  satisfying the following conditions.  \smallskip
 
 {\sc Assumption (B)}
 For any $f\in\mathcal{F}_1$ and $\varthetab_{\zerob}$, under ${\rm P}_{\varthetab_{\zerob};f}^{(n)}$, as $n\to\infty$, 
  $\hat\thetab\n$\linebreak and~$\hat\Sigmab\n$  {\it (i)} are root-$n$ consistent:  $n^{1/2}(\hat\thetab\n-\thetab)$ and $n^{1/2}(\hat\Sigmab\n-\Sigmab)$ are~$O_{\rm P}(1)$, and 
 {\it (ii)} are locally asymptotically discrete: 
  the number of possible values of~$\hat\thetab\n$ and ${\rm vech}\hat\Sigmab\n$ in any sequence of $O(n^{-1/2})$ balls centered around $\thetab$ and ${\rm vech}\Sigmab$, respectively, 
 is  uniformly bounded as $n\rightarrow\infty$.\medskip

This assumption, in combination with Lemma 4.4 of \citet{kreiss1987adaptive}, entails 
\begin{eqnarray} \nonumber
\Deltab_f(\hat\thetab\n\!,\hat\Sigmab\n\!\!,\zerob)&-& \Deltab_f(\thetab,\Sigmab,\zerob)\\ \label{aslin2}
&=& -\Gammab_f\big(\varthetab_{\pmb{0}})n^{1/2}((\hat\thetab^{(n)\prime}\!\!,({\rm vech}\hat\Sigmab\n)'\!\!,\zerob'\big)'-\varthetab_{\zerob}) + o_{\rm P}(1)
\end{eqnarray}
under ${\rm P}_{\varthetab_{\zerob};f}^{(n)}$ as $n \to \infty$.
%
It should be noted that Assumption B{\it (ii)} is a purely technical requirement, with little practical implications (for fixed sample size, any estimator indeed can be considered part of a locally asymptotically discrete sequence: see \citet{yang2000asymptotics}). 

In practice, it is   desirable   to restrict to affine-equivariant estimators:  we will assume that $\hat\thetab\n$ and $\hat\Sigmab\n$ also satisfy 
$$\hat\thetab\n(\Ab\Xb_1+\bb,\ldots,\Ab\Xb_n+\bb)=\Ab\hat\thetab\n(\Xb_1,\ldots,\Xb_n)+\bb\vspace{-2mm}$$ 
and \vspace{-2mm}
$$\hat\Sigmab\n(\Ab\Xb_1+\bb,\ldots,\Ab\Xb_n+\bb)=\Ab\hat\Sigmab\n(\Xb_1,\ldots,\Xb_n)\Ab'$$
 for any $d\times d$ matrix $\Ab$ and any $d$-vector $\bb$. Under this natural requirement,  our test statistics will enjoy affine-invariance. In the sequel, the lighter notation $\hat\thetab$,  $\hat\Sigmab$ will be adopted.

We conclude this section on ULAN by  noting the block-diagonal structure of the Fisher information matrix, implying that the $\Sigmab$- and $(\thetab,\lambdab)$-parts of the central sequence are asymptotically independent. 

\section{Optimal parametric and semiparametric tests: 
  specified~$\thetab$}\label{sec:fix}

Fix $\thetab\in\R^d$.  ULAN  and  the convergence of local sequences of experiments to a Gaussian  shift experiment  imply that a locally asymptotically optimal parametric test for  $\mathcal{H}_{0;f,\thetab}$ against $\mathcal{H}_{1;f,\Pi,\thetab}$ can  be based on a quadratic form involving  the $\lambdab$-part $\Deltab_{3}(\varthetab_{\pmb{0}})$ of the central sequence. Of course, the nuisance scatter parameter $\Sigmab$ needs to be estimated. The block-diagonal structure of the Fisher information matrix, combined with~\eqref{aslin2}   allows for substituting,  without any loss of power, any    $\hat\Sigmab$ satisfying Assumption~(B) for the unknown~$\Sigmab$.  Thus, unlike Rao score/Lagrange multiplier tests or likelihood ratio tests, where $\hat\Sigmab$ has to be the MLE, we can accommodate various estimators and privilege computational convenience or robustness, or avoid higher-order moment assumptions. In the sequel, we are opting for \citet{tyler1987distribution}'s estimator of scatter (shape). 
Denote by $\mathbf{T}$ the unique (for~$n>d(d-1)$) $d\times d$ upper-triangular matrix with positive diagonal elements and determinant equal to one satisfying
\begin{displaymath}
\frac{1}{n}\sum_{i=1}^n\left(\frac{\mathbf{T}(\Xb_i-\thetab)}{\|\mathbf{T}(\Xb_i-\thetab)\|}\right)\left(\frac{\mathbf{T}(\Xb_i-\thetab)}{\|\mathbf{T}(\Xb_i-\thetab)\|}\right)'=\frac{1}{d}\mathbf{I}_d.
\end{displaymath}
This matrix $\mathbf{T}$ is such  that the  covariance structure of 
\begin{displaymath}
\left(\frac{\mathbf{T}(\Xb_1-\thetab)}{\|\mathbf{T}(\Xb_1-\thetab)\|},\ldots,\frac{\mathbf{T}(\Xb_n-\thetab)}{\|\mathbf{T}(\Xb_n-\thetab)\|}\right)\vspace{-1mm}
\end{displaymath}
is   that of an i.i.d.\ sample with uniform distribution over the unit sphere in~$\R^d$. Tyler's  estimator of shape is then   $(\mathbf{T}\mathbf{T}')^{-1}$ which  we turn into a scatter estimator in accordance with the integration condition in the definition of~$\mathcal{F}_1$.


 Another potential estimator of $\Sigmab$ is the {\it minimum covariance determinant} (MCD) estimator 
   (\citep{rousseeuw1984least},\citep{rousseeuw1999fast}). Both Tyler's and the MCD estimator are affine-invariant. 

Letting $\hat\varthetab_0:=(\thetab',(\vech\hat\Sigmab)',\pmb{0}')'$ for some 
 estimator $\hat\Sigmab$ satisfying Assumption~(B), denote by $\phi_{\thetab;f}^{(n)}$ the test rejecting the null hypothesis $\mathcal{H}_{0;f,\thetab}$ 
  whenever
$$
Q_{\thetab;f}^{(n)}:=(\Deltab_{3}(\hat\varthetab_0))'(\Gammab_{f;33}(\varthetab_0))^{-1}\Deltab_{3}(\hat\varthetab_0)
$$
exceeds   the $\alpha$-upper quantile $\chi^2_{d;1-\alpha}$ of the chi-squared distribution with~$d$ degrees of freedom. This asymptotic null distribution  easily follows from the asymptotic normality of $\Deltab_{3}(\varthetab_0)$ and the fact that $\Deltab_{3}(\hat\varthetab_0)-\Deltab_3(\varthetab_0)$ is~$o_{\rm P}(1)$ under $\mathcal{H}_{0;f,\thetab}$ as $n\rightarrow\infty$. The  test $\phi_{\thetab;f}^{(n)}$ is locally and asymptotically optimal 
 for $\mathcal{H}_{0;f,\thetab}$ against $\mathcal{H}_{1;f,\Pi,\thetab}$ (see Theorem~\ref{asympt} for  its precise optimality properties). 

  Elementary algebra yields\vspace{-1mm} 
$$
Q_{\thetab;f}^{(n)}=n (\bar{\Xb}-\thetab_0)'\hat\Sigmab^{-1}(\bar{\Xb}-\thetab_0)=:Q_{\thetab}^{(n)}.
$$
This expression is particularly striking, as it does \emph{not} depend on the underlying radial density $f$.  In other words, every parametric specified-$f$ experiment leads to the same optimal test statistic  $Q_{\thetab}^{(n)}$, so that $\phi_{\thetab}^{(n)}:=\phi_{\thetab;f}^{(n)}$ is {\it uniformly} (in~$f$) optimal in the semiparametric unspecified-$f$ experiment. 
This is an extremely rare feature. Another remarkable fact is that the skewing function~$\Pi$ plays no role in $Q_{\thetab}^{(n)}$, which means that optimality holds uniformly against {\it all} skew-elliptical alternatives. 
Finally, the alert reader has noticed the familiar form of $Q_{\thetab}^{(n)}$, which is nothing else but  the classical Hotelling test statistic for location. Optimal testing for ellipticity with specified location thus, somewhat disappointingly, mostly boils down to testing for location.
%

  The following theorem summarizes the properties of $\phi_{\thetab}^{(n)}$. 
\begin{thm}\label{asympt}
Let $f\in\mathcal{F}_1$ and suppose that Assumptions (A1), (A2), and~(B) hold, and that the skewing function $\Pi$ is continuously differentiable  at~0, with~$\dot{\Pi}(0)\neq0$.  Then,

\noindent(i) under  $\mathcal{H}_{0;\thetab}$, $Q_{\thetab}^{(n)}\stackrel{\mathcal{D}}{\rightarrow} \chi^2_d$ as $n\to\infty$, so that 
$\phi^{(n)}_{\thetab}$ has asymptotic level $\alpha$;

\noindent(ii) under $\bigcup_{\,\Sigmab\in\mathcal{S}_d} {\rm P}^{(n)}_{\thetab,\Sigmab,n^{-1/2}\taub^{(n)}_3;g,\Pi}$  with $g\in\mathcal{F}_1$, $Q_{\thetab}^{(n)}$ is asymptotically non-central chi-square  with $d$ degrees of freedom and non-centrality parame\-ter~$4(\dot{\Pi}(0))^2\taub_3'\taub_3$, where  $\taub_3=\lim_{n\rightarrow\infty}\taub^{(n)}_3$\footnote{Here and in the sequel, several asymptotic results are established for sequences of perturbations of the form $n^{-1/2}\taub^{(n)}_3$ such that $\taub^{(n)}_3$ converges to $\taub_3$. Clearly, since $\taub^{(n)}_3$ is bounded, converging subsequences always exist; the asymptotic statement then holds along any such subsequence. This is tacitly assumed below whenever defining $\taub_3$ as the limit of a sequence  $\taub^{(n)}_3$.};

\noindent(iii) 
the test~$\phi^{(n)}_{\thetab}$ is locally and asymptotically maximin, at asymptotic level~$\alpha$, for testing $\mathcal{H}_{0;\thetab}$ against $\mathcal{H}_{1;\Pi,\thetab}=\bigcup_{\,f\in\mathcal{F}_1,\,\Sigmab\in\mathcal{S}_d,\,\lambdab\in\R^d\setminus\{\zerob\}} {\rm P}^{(n)}_{\thetab,\,\Sigmab,\,\lambdab;f,\Pi}$. The test is thus uniformly (in $f$) optimal against any type of generalized skew-elliptical alternative as defined in~\eqref{skewell}.
\end{thm}
\vspace{-1mm}

The proof is provided in  Appendix~B. The explicit expression 
$$
1-F_{\chi^{'2}_d}(\chi^2_{d;1-\alpha},4(\dot{\Pi}(0))^2\taub_3'\taub_3) = Q_{d/2}\left(2|\dot{\Pi}(0)|({\taub_3'\taub_3})^{1/2},({\chi^2_{d;1-\alpha}})^{1/2}\right)\vspace{-1mm}
$$
of the asymptotic power of $\phi^{(n)}_{\thetab}$ against   local alternatives of the form~$\bigcup_{\,\Sigmab\in\mathcal{S}_d} {\rm P}^{(n)}_{\thetab,\Sigmab,n^{-1/2}\taub^{(n)}_3;g,\Pi}$ readily follows from part~{\it (ii)} of the theorem ($F_{\chi^{'2}_d}$ stands for the   distribution function of the non-central chi-square distribution with $d$ degrees of freedom,   $Q_M(\cdot,\cdot)$ for the {\it Marcum Q-function}).

\section{Optimal parametric and semiparametric tests: unspecified $\thetab$}\label{sec:notfix}

In some applications, maintaining a specified   value of $\thetab$ (often, $\thetab = \zerob$) under the alternative does make sense. The test described in Theorem~\ref{asympt} then is a genuine  test of ellipticity. In most cases, however, that assumption of a specified center is impossible or unrealistic---or just unclear: what is the ``center'' of an asymmetric distribution? 
 The same test then no longer qualifies as a test of ellipticity. Moreover, as shown in Section~\ref{numsec},  the impacts    of location shift and  non-ellipticity may cancel each other, with the consequence that obviously non-elliptical   shifted distributions remain completely undetected (see Sections~\ref{pitsec} and~\ref{empirsec} for numerical evidence). 
Therefore, let us consider  the case of an unspecified $\thetab$. \smallskip

Instances of estimators of $\thetab$ that satisfy Assumption~(B) and turn out to be useful in this section are   the {\it spatial median} of \citet{MO95} or the (fast) MCD-based location estimator (\citet{rousseeuw1999fast}). Again, we shall first construct Le Cam efficient parametric tests (Section~\ref{optpar}) and then turn them into semiparametrically efficient  tests (Section~\ref{optsemi}). 

Inspection of the 
Fisher information matrix \eqref{Fishinfo} reveals that the scores for location and skewness are not asymptotically independent. Estimating the unknown location thus has a cost in terms of power against ellipticity. The family of generalized skew-elliptical distributions, moreover, is infamous for yielding   singular Fisher information matrices in the vicinity of symmetry, which is precisely the situation we are interested in. In presence of such a singularity, the scores for skewness and location are perfectly colinear, with the consequence that the corresponding $\alpha$-level optimal test for symmetry is  the trivial test~$\phi=\alpha$. 
Fortunately, this extreme situation only occurs at the multinormal distribution (\cite{ley2010singularity}, \cite{hallin2012skew}, \cite{hallin2014skew}). Testing for multinormality against generalized skew-normality thus requires a special treatment (reparametrization and ULAN with slower contiguity rates), which is beyond the scope of this paper. 

\subsection{Optimal parametric tests: unspecified  $\thetab$}\label{optpar}

Fix a radial density $f\in~\!\mathcal{F}_1$ that is not Gaussian. The impact on  the central sequence for skewness~$\Deltab_{3}(\varthetab_0)$ of a root-$n$ perturbation of $\thetab$ is classically neutralized by projecting~$\Deltab_{3}(\varthetab_0)$ onto the subspace orthogonal to  $\Deltab_{f;1}(\varthetab_0)$ in the metric of the information matrix, yielding 
 the \emph{$f$-efficient central sequence for skewness}
\begin{align*}
    {\Deltab}^\dagger_{f;3}(\varthetab_0) := \Deltab_{3}(\varthetab_0) - \Gammab_{f;13}(\varthetab_0) \Gammab_{f;11}^{-1}(\varthetab_0) \Deltab_{f;1}(\varthetab_0).
\end{align*}
Clearly, this new central sequence remains orthogonal to $\Deltab_{f;2}(\varthetab_0)$. This orthogonality to $\Deltab_{f;1}$ and $\Deltab_{f;2}$, combined with \eqref{aslin2},  allows us to replace the unknown parameters $\Sigmab$ and $\thetab$ with any   consistent estimators $\hat{\Sigmab}$ and $\hat{\thetab}$ satisfying Assumption~(B)  without altering the asymptotic behavior of~$ {\Deltab}^\dagger_{f;3}$ under the null and under local alternatives.  
Under  $\mathcal{H}_{0;f}$, ${\Deltab}^\dagger_{f;3}(\varthetab_0)$, hence also~${\Deltab}^\dagger_{f;3}(\hat\varthetab_0)$,   is asymptotically normal with
mean zero and covariance (the~{\it $f$-efficient Fisher information for skewness})
\begin{align*}
    \Gammab^\dagger_{f;33}(\varthetab_0) :=
\Gammab_{f;33}(\varthetab_0)-
\Gammab_{f;13}(\varthetab_0) \Gammab_{f;11}^{-1}(\varthetab_0) \Gammab_{f;13}(\varthetab_0).
\end{align*}
Note that this matrix would be the zero matrix if $f$ were Gaussian. The resulting optimal $f$-parametric test statistic  then is of the form
\begin{eqnarray*}
    Q_{f}^{(n)}&:=&({\Deltab}^\dagger_{f;3}(\hat\varthetab_0))'\Big(\Gammab^\dagger_{f;33}(\hat\varthetab_0)\Big)^{-1}\Deltab^\dagger_{f;3}(\hat\varthetab_0)
    \\
    & =&   \frac{{\mathcal{I}}_{d,f}}{({\mathcal{I}}_{d,f}-d)} \frac{1}{n}\sum_{i,j=1}^n\Big[\din(\hat\thetab,\hat\Sigmab)-\frac{d}{{\mathcal{I}}_{d,f}}\varphi_f(\din(\hat\thetab,\hat\Sigmab))\Big]
    \\
    &&\times
    \Big[\djn(\hat\thetab,\hat\Sigmab)-\frac{d}{{\mathcal{I}}_{d,f}}\varphi_f(\djn(\hat\thetab,\hat\Sigmab))\Big]\Ubin(\hat\thetab,\hat\Sigmab))'\Ubjn(\hat\thetab,\hat\Sigmab),
\end{eqnarray*}
and the corresponding test $\phi_f^{(n)}$ rejects $\mathcal{H}_{0;f}$ at asymptotic level $\alpha$ whenever~$Q_{f}^{(n)}$ exceeds the chi-square quantile $\chi^2_{d;1-\alpha}$. The next theorem, the proof of which we give in Appendix C, summarizes the asymptotic   properties of this~test.

\begin{thm} \label{asympt2}
Let $f\in \mathcal{F}_1$ and suppose that Assumptions (A1), (A2), and~(B) hold, and that the skewing function $\Pi$ is continuously differentiable at $0$, with derivative $\dot{\Pi}(0)\neq0$.  Then,

\noindent(i) under  $\mathcal{H}_{0;f}$, $Q_{f}^{(n)}\stackrel{\mathcal{D}}{\rightarrow} \chi^2_d$ as $n\to\infty$, so that~$\phi^{(n)}_{f}$ has asymptotic level $\alpha$;

\noindent(ii) under $\bigcup_{\,\thetab\in\R^d}\bigcup_{\,\Sigmab\in\mathcal{S}_d} {\rm P}^{(n)}_{\thetab,\Sigmab,n^{-1/2}\taub^{(n)}_3;f,\Pi}$, $Q_{f}^{(n)}$ is asymptotically non-cen\-tral chi-square   with $d$ degrees of freedom and non-centrality para\-me\-ter~$4(\dot{\Pi}(0))^2\big( ({\mathcal{I}_{d,f} - d})/{\mathcal{I}_{d,f}}\big) \taub_3'\taub_3$, where  $\taub_3=\lim_{n\rightarrow\infty}\taub^{(n)}_3$

\noindent(iii) 
the test~$\phi^{(n)}_{f}$ is locally and asymptotically maximin, at asymptotic level~$\alpha$, for $\mathcal{H}_{0;f}$ against $\mathcal{H}_{1;f,\Pi}=\bigcup_{\,\thetab\in\R^d,\Sigmab\in\mathcal{S}_d,\lambdab\in\R^d\setminus\{\zerob\}} {\rm P}^{(n)}_{\thetab,\,\Sigmab,\,\lambdab;f,\Pi}$. The test is thus optimal against any type of generalized skew-$f$ alternative.\vspace{-1mm}
\end{thm}
Summing up, the test $\phi^{(n)}_{f}$ is (parametrically) optimal against any type of generalized skew-$f$ alternative ($f$ specified).

\subsection{Optimal semiparametric tests: unspecified $\thetab$}\label{optsemi}

Consider now the general null hypothesis $\mathcal{H}_0$ of elliptical symmetry with unspecified center $\thetab$. Since the central sequence for skewness $\Deltab_{3}(\varthetab_0)$  does not depend on the actual radial density, the ideal test for the  case of unspecified $f$ and $\thetab$ should be based on $\Deltab_{3}(\varthetab_0)$. But $\Deltab_{3}(\varthetab_0)$ also depends on $\thetab$ and $\Sigmab$, which therefore have to be replaced with estimators $\hat\thetab$ and $\hat\Sigmab$ and, unfortunately, the impact of that substitution does depend on the actual radial density (denote it as~$g$). 

Let $\hat\thetab$ and $\hat\Sigmab$ satisfy 
 Assumption~(B). The asymptotic linearity property~(note that \eqref{aslin2} applies under any  ${\rm P}^{(n)}_{\thetab,\,\Sigmab,\,\zerob;g}$ thanks to the fact that~$\Deltab_{3}(\varthetab_0)$  does not depend on $g$)
  yields,  under ${\rm P}^{(n)}_{\thetab,\,\Sigmab,\,\zerob;g}$  as $n\rightarrow\infty$, 
$$
\Deltab_{3}(\hat\thetab,\hat\Sigmab,\zerob)-\Deltab_{3}(\thetab,\Sigmab,\zerob)=-\text{Cov}_{g}\left[\Deltab_{3}(\varthetab_0),\Deltab_{g;1}(\varthetab_0) \right] n^{1/2}(\hat\thetab-\thetab)+o_{\rm P}(1)
$$
where $\text{Cov}_g\left[ \Deltab_{3}(\varthetab_0), \Deltab_{g;1}(\varthetab_0) \right] = 2\dot{\Pi}(0)\Sigmab^{-1/2}$.  This is a non-zero quantity    the projection of the previous section cannot cancel out for all $g$. Therefore, a ``deeper projection'' is required  to obtain an $f$-efficient central sequence that is orthogonal, under ${\rm P}^{(n)}_{\thetab,\,\Sigmab,\,\zerob;g}$, to the $g$-based central sequen\-ce~$\Deltab_{g;1}(\varthetab_0)$,   for any $g$. This deeper projection is taken care of by 
   \begin{align*}
   \displaystyle{\Deltab}^\ddagger_{fg;3}(\varthetab_0) := \Deltab_{3}(\varthetab_0)-2\dot{\Pi}(0)\Sigmab^{-1/2} \left[\text{Cov}_{g}\left[ \Deltab_{f;1}(\varthetab_0), \Deltab_{g;1}(\varthetab_0) \right]\right]^{-1}   \Deltab_{f;1}(\varthetab_0)
    \end{align*}
which, unfortunately, depends on the unspecified $g$ again. Simple  algebra yields
 $\text{Cov}_g\left[ \Deltab_{f;1}(\varthetab_0), \Deltab_{g;1}(\varthetab_0) \right] = \frac{1}{d}\mathcal{K}_{d,f,g}\Sigmab^{-1}$,  with\vspace{-1mm}
 $$
  \mathcal{K}_{d,f,g}:= {\displaystyle\int}_{0}^\infty\displaystyle \left(\varphi_f^\prime (r) + \frac{d-1}{r}\varphi_f(r)\right)\frac{1}{\mu_{d-1;g}} r^{d-1}g(r)dr
 \vspace{-1mm} $$
where we denote by   $\varphi_f^\prime$ the weak derivative of $r\mapsto \varphi_f(r)$ coinciding, in case~$\varphi_f$ is differentiable, with the usual derivative of $r\mapsto\varphi_f(r)$. 

The existence of this latter quantity, however, requires a slight reinforcement of the assumptions {on the reference radial densities $f$ and the actual radial density $g$}. 

{\sc Assumption (A3)}
The mapping $r\mapsto f^{1/2}(r)$ belongs to~$W^{2,2}(\mathbb{R}_0,\mu_{d-1})$, $0\neq \left| 
{\displaystyle\int}_{0}^\infty\displaystyle \left(\varphi_f^\prime (r) + \frac{d-1}{r}\varphi_f(r)\right) r^{d-1}f(r)dr\right|<\infty$, and~$\displaystyle\int_0^\infty (\varphi (r))^{2+\epsilon}r^{d-1}f(r) dr <\infty$ for some $\epsilon >0$.\smallskip

It follows from the definition 
 that, for any $f\in\mathcal{F}_1$ satisfying Assumptions (A1-A3), there exists a class  of densities  
\begin{eqnarray*}
\mathcal{F}_{1;f}&:=&\left\{g\in\mathcal{F}_{1}:0\neq \left|\int_{0}^\infty\displaystyle \left(\varphi_f'(r) + \frac{d-1}{r}\varphi_f(r)\right) r^{d-1}g(r)dr\right|<\infty\,\right.\\
&&\quad\quad\quad\quad\quad\left.\mbox{and}\, \int_0^\infty(\varphi_f(r))^{2+\epsilon_g}r^{d-1}g(r) dr<\infty\,\mbox{for some}\ \epsilon_g>0\right\}\vspace{-1mm}
\end{eqnarray*}
such that, for $g\in \mathcal{F}_{1;f}$, $\mathcal{K}_{d,f,g}$, hence ${\Deltab}^\ddagger_{fg;3}$,  are well defined. Clearly, under Assumptions (A1) and~(A3),  $f$ itself belongs to~$\mathcal{F}_{1;f}$.  The resulting projected central sequence is  
   \begin{align*}
   \displaystyle{\Deltab}^\ddagger_{fg;3}(\varthetab_0) =  2n^{-1/2}\dot{\Pi}(0)\sum_{i=1}^n\left[\din(\thetab,\Sigmab)-\frac{d}{{\mathcal{K}}_{d,f,g}}\varphi_f(\din(\thetab,\Sigmab))\right]\Ubin(\thetab,\Sigmab).
   \end{align*}
This, through $\mathcal{K}_{d,f,g}$, still depends on the unknown $g$. But $\mathcal{K}_{d,f,g}$  can be estimated via 
 \begin{align*}
          \widehat{\mathcal{K}}_{d,f}(\thetab,\Sigmab) := \frac{1}{n}\sum_{i=1}^n \left[ \displaystyle \varphi_f'(\din(\thetab,\Sigmab)) + \frac{d-1}{\din(\thetab,\Sigmab)}\varphi_f(\din(\thetab,\Sigmab)) \right],
      \end{align*}
hence, \emph{in fine}, just as for the entire test statistic, by $\widehat{\mathcal{K}}_{d,f}(\hat\thetab,\hat\Sigmab)$ with $\hat\thetab$ and $\hat\Sigmab$  satisfying Assumption (B). The following lemma   establishes the consistency   of $ \widehat{\mathcal{K}}_{d,f}(\hat\thetab,\hat\Sigmab)$ as an estimator of $\mathcal{K}_{d,f,g}$.
\begin{lem}\label{Lemma1}
Let $f\in \mathcal{F}_1$ and suppose that Assumptions (A1-A3) and~(B) hold. Then,  for any $g\in \mathcal{F}_{1;f}$, $ \widehat{\mathcal{K}}_{d,f}(\hat\thetab,\hat\Sigmab)-\mathcal{K}_{d,f,g}=o_{\rm P}(1)$ as $n\rightarrow\infty$ under~${\rm P}^{(n)}_{\thetab,\,\Sigmab,\,\zerob;g}$.
\end{lem}
The proof is provided in Appendix~C.

With this estimator of $\mathcal{K}_{d,f,g}$, the  efficient central sequence for skewness takes  the final form      \begin{align*}
   \displaystyle {\Deltab}^\ddagger_{f;3}(\hat\varthetab_0) =  2n^{-1/2}\dot{\Pi}(0)\sum_{i=1}^n\left[\din(\hat\thetab,\hat\Sigma)-\frac{d}{ \widehat{\mathcal{K}}_{d,f}(\hat\thetab,\hat\Sigmab)}\varphi_f(\din(\hat\thetab,\hat\Sigmab))\right]\Ubin(\hat\thetab,\hat\Sigmab).
   \end{align*}
The corresponding test
 $\phi^{\ddagger (n)}_f$ rejects 
   $\mathcal{H}_0$ at asymptotic level $\alpha$ whenever the test statistic 
$  Q^{\ddagger (n)}_f:= ({\Deltab}^\ddagger_{f;3}(\hat\varthetab_0))' (\widehat{\Gammab}^\ddagger_{f}(\hat\varthetab_0))^{-1}{\Deltab}^\ddagger_{f;3}(\hat\varthetab_0)$, 
    with 
 $$\widehat{\Gammab}^\ddagger_{f}(\hat\varthetab_0):=\frac{{4(\dot{\Pi}(0))^2}}{nd}\sum_{i=1}^n\left[\din(\hat\thetab,\hat\Sigmab)-\frac{d}{ \widehat{\mathcal{K}}_{d,f}(\hat\thetab,\hat\Sigmab)}\varphi_f(\din(\hat\thetab,\hat\Sigmab))\right]^2\Ik$$
%
exceeds the chi-square quantile $\chi^2_{d;1-\alpha}$. The asymptotic distribution of $Q^{\ddagger (n)}_f$ under ${\rm P}^{(n)}_{\thetab,\,\Sigmab,\,\zerob;g}$ for any $g\in\mathcal{F}_{1;f}$ and its optimality properties are formally established in Theorem~\ref{asympt3}.
For the sake of exposition, we first establish the following lemma   (see Appendix~C for a proof).
\begin{lem}\label{Lemma2}
Let $f\in \mathcal{F}_1$ and suppose that Assumptions (A1-A3) and (B) hold. Then,
\begin{enumerate}
\item[(i)] ${\Deltab}^\ddagger_{f;3}(\hat\varthetab_0)-{\Deltab}^\ddagger_{fg;3}(\varthetab_0)=o_{\rm P}(1)$ and  
\item[(ii)] $\widehat{\Gammab}^\ddagger_{f}(\hat\varthetab_0)-\Gammab^\ddagger_{f}(\varthetab_0)=o_{\rm P}(1)$\vspace{-1mm}
\end{enumerate}
 as~$n\rightarrow\infty$ under ${\rm P}^{(n)}_{\thetab,\,\Sigmab,\,\zerob;g}$ for any $g\in \mathcal{F}_{1;f}$, where\vspace{-1mm}  
$$
{\Gammab}^\ddagger_{f}(\varthetab_0):=\frac{{4(\dot{\Pi}(0))^2}}{nd}\sum_{i=1}^n\left[\din(\thetab,\Sigmab)-\frac{d}{{\mathcal{K}}_{d,f,g}}\varphi_f(\din(\thetab,\Sigmab))\right]^2\Ik.\vspace{-1mm}
$$
\end{lem}
With this result in hand, we finally can state the announced asymptotic results about $\phi^{\ddagger (n)}_f$ and $Q^{\ddagger (n)}_f$.\vspace{-1mm}

\begin{thm} \label{asympt3}
Let $f\in \mathcal{F}_1$ and suppose that Assumptions (A1-A3) and~(B) hold,  and that the skewing function $\Pi$ is continuously differentiable at $0$, with $\dot{\Pi}(0)\neq0$.  Then,

\noindent(i) under   $\bigcup_{\,g\in\mathcal{F}_{1;f},\,\thetab\in\Rd\!,\,\Sigmab\in\mathcal{S}_d} {\rm P}^{(n)}_{\thetab,\,\Sigmab,\,\zerob;g}$, the test statistic $Q^{\ddagger (n)}_f$ is asymptotically~$\chi^2_d$ as $n\to\infty$, so that  the  test~$\phi^{\ddagger (n)}_f$ has asymptotic level $\alpha$;

\noindent(ii) under $\bigcup_{\,\thetab\in\R^d}\bigcup_{\,\Sigmab\in\mathcal{S}_d} {\rm P}^{(n)}_{\thetab,\Sigmab,n^{-1/2}\taub^{(n)}_3;g,\Pi}$ with  $g\in \mathcal{F}_{1;f}$, $Q^{\ddagger (n)}_f$ is~asymp\-tot\-i\-cal\-ly non-central chi-square  with  $d$ degrees of freedom and non-centrality parameter $\displaystyle 4(\dot{\Pi}(0))^2d\gamma_{d,f,g}^{-1}{(1- {\alpha_{d,f,g}}/{{\mathcal{K}_{d,f,g}}})^2}\taub_3'\taub_3$, where $\taub_3=\lim_{n\rightarrow\infty}\taub^{(n)}_3$, $
\alpha_{d,f,g}:=\,\frac{1}{\mu_{d-1;g}}\int_{0}^{\infty}\displaystyle r \varphi_f(r)r^{d-1} g(r)dr
$, and\vspace{-2mm}
$$
\gamma_{d,f,g}:=\,\frac{1}{\mu_{d-1;g}}\int_{0}^{\infty}\displaystyle \left[r - \frac{d}{{\mathcal{K}_{d,f,g}}}\varphi_f(r)\right]^2 r^{d-1}g(r)dr
.\vspace{-2mm}$$

\noindent(iii) 
the test~$\phi^{\ddagger (n)}_f$ is locally and asymptotically maximin, at asymptotic level~$\alpha$, when testing $\bigcup_{\,g\in\mathcal{F}_{1;f},\thetab\in\Rd\!,\Sigmab\in\mathcal{S}_d} {\rm P}^{(n)}_{\thetab,\,\Sigmab,\,\zerob;g}$ against alternatives of the form~$\mathcal{H}_{1;f,\Pi}=\bigcup_{\,\thetab\in\R^d,\,\Sigmab\in\mathcal{S}_d,\,\lambdab\in\R^d\setminus\{\zerob\}} {\rm P}^{(n)}_{\thetab,\,\Sigmab,\,\lambdab;f,\Pi}\vspace{1mm}$, irrespective of $\Pi$.
\end{thm}
Part~{\it (i)}  of this Theorem easily follows from Lemma~\ref{Lemma2}. The rest of the proof follows along the same lines as the proofs of Theorems~\ref{asympt} and~\ref{asympt2}; details are   left to the reader. Note that the finiteness of $\gamma_{d,f,g}$ follows from our assumptions on $g$.\smallskip

The test $\phi^{\ddagger (n)}_f$ thus is valid under any $g\in\mathcal{F}_{1;f}$---the entire nonparametric hypothesis of elliptical symmetry with unspecified center---and uniformly optimal against {\it any}  type of generalized skew-$f$ alternative. For each  radial density $f$ satisfying Assumptions (A1-A3), we thus get such a test $\phi^{\ddagger (n)}_f$. These tests are the main contribution 
%
  of this paper, and achieve all our objectives: they have a simple asymptotic chi-squared distribution under the  null hypothesis of ellipticity, they are affine-invariant (this follows directly from the affine-invariance of $\din(\hat\thetab,\hat\Sigmab)$ and $\Ubin(\hat\thetab,\hat\Sigmab)$), computationally fast, have a simple and intuitive form, only require finite moments of order 2, and offer much flexibility in the choice of the radial density $f$ at which  optimality is achieved  (recall that a Gaussian $f$ is excluded, though). 

The choice of $f$ can be guided by asymptotic relative efficiency profiles, which we now provide for various  choices of $f$.

\subsection{Asymptotic Relative Efficiencies}\label{ARE}
In this section, we compute  Asymptotic Relative Efficiencies (AREs) for $\phi^{\ddagger (n)}_f$ with respect to   the pseudo-Gaussian test of \citep{Cassart07} as a common benchmark. 

Define $m_{k}^{(n)}(\thetab,\Sigmab) := n^{-1}\sum_{i=1}^{n}(d_{i}(\thetab,\Sigmab))^k$ and \vspace{-1mm}
\begin{align}\nonumber\Sb^{\Ub}_i(\thetab,\Sigmab) := ((\Ub_{i1}(\thetab,\Sigmab))^2 \text{sign}&(\Ub_{i1}(\thetab,\Sigmab)), \dots
\\ 
\dots&, (\Ub_{id}(\thetab,\Sigmab))^2 \text{sign}(\Ub_{id}(\thetab,\Sigmab)))^\prime . \vspace{-1mm}
\label{Sdef}\end{align}
When the location $\thetab$ 
 is unspecified, the  Gaussian efficient central sequence for  Cassart's Fechner-asymmetry model is  
$$
\Deltab_{{{\cal G}}}(\varthetab_0)=  n^{-1/2}\sum_{i=1}^{n}d_i(\thetab,\Sigmab)(c_d(d+1)m_{1}^{(n)}(\thetab,\Sigmab)\Ub_i(\thetab,\Sigmab)-d_i(\thetab,\Sigmab)\Sb^{\Ub}_i(\thetab,\Sigmab))
$$
where $c_d = 4\Gamma(d/2)/((d^2-1)\sqrt{\pi}\Gamma(\frac{d-1}{2})$, with  Fisher information matrix under radial density $g$ (note that $m_{k}^{(n)}(\thetab,\Sigmab)$ converges to $\frac{\mu_{d+k-1;g}}{\mu_{d-1;g}}$ under $g$)
$$\Gammab_{{{\cal G}}}(\varthetab_0) := \left(\frac{3}{d(d+2)}\frac{\mu_{d+3;g}}{\mu_{d-1;g}} - 2c_{d}^2(d+1)\frac{\mu_{d;g}\mu_{d+2;g}}{(\mu_{d-1;g})^2} + c_{d}^{2}\frac{(d+1)^2}{d}\frac{\mu_{d;g}^{2}\mu_{d+1;g}}{(\mu_{d-1;g})^3}\right)\Ik .$$  

The expectation of $\Deltab_{{{\cal G}}}(\varthetab_0)$ remains $\zerob$, and the asymptotic normality with covariance $\Gammab_{{{\cal G}}}(\varthetab_0)$ holds, under any 
 $g$ with finite fourth-order moment, that is, under 
 $
g\in\mathcal{F}_{{p{\cal G}}}:=\left\{f \in \mathcal{F}_1: \mu_{d+3;f}=\int_{\R^+}r^{d+3}f(r)dr<\infty \right\}.
$

The Gaussian test based on $\Deltab_{{{\cal G}}}(\varthetab_0)$ thus can be used as a 
 pseudo-Gaussian test: denote it as $\phi_{{p{\cal G}}}^{(n)}$. That test rejects the null hypothesis~$\bigcup_{\,g\in\mathcal{F}_{{p{\cal G}}},\,\thetab\in\Rd\!,\,\Sigmab\in\mathcal{S}_d} {\rm P}^{(n)}_{\thetab,\,\Sigmab,\,\zerob;g}$ of elliptical symmetry with unspecified~$g$ and~$\thetab$  at asymptotic level $\alpha$ whenever the test statistic (with $\hat\thetab$ and $\hat\Sigmab$ satisfying Assumption (B)) 
$
Q_{{p{\cal G}}}^{(n)} := (\Deltab_{{{\cal G}}}(\hat\varthetab_0))'(\Gammab_{{{\cal G}}}(\hat\varthetab_0))^{-1}\Deltab_{{{\cal G}}}(\hat\varthetab_0)
$ 
exceeds $\chi^2_{d;1-\alpha}$. We refer   to Chapter 3 of \cite{Cassart07} for formal details.

In order to compute AREs   with respect to $\phi_{{p{\cal G}}}^{(n)}$, we need its asymptotic distribution   under the  local skew-elliptical alternatives considered in this paper. This is the purpose of the following result, the proof of which is similar to those of Theorems~\ref{asympt} and~\ref{asympt2} and is left to the reader.

\begin{thm}\label{propo}
Suppose that Assumptions (A1), (A2), and (B) hold, and that the skewing function $\Pi$ is continuously differentiable at $0$ with~$\dot{\Pi}(0)\neq0$.   Then,

\noindent(i) under  $\bigcup_{\,g\in\mathcal{F}_{{p{\cal G}}},\,\thetab\in\Rd\!,\,\Sigmab\in\mathcal{S}_d} {\rm P}^{(n)}_{\thetab,\,\Sigmab,\,\zerob;g}\vspace{1mm}$, $Q_{{p{\cal G}}}^{(n)}$ is asymptotically $\chi^2_d$ as $n\to\infty$, so that~$\phi_{{p{\cal G}}}^{(n)}$ has asymptotic level $\alpha$;

\noindent(ii)
under $\bigcup_{\,\thetab\in\R^d}\cup_{\Sigmab\in\mathcal{S}_d} {\rm P}^{(n)}_{\thetab,\Sigmab,n^{-1/2}\taub^{(n)}_3;g,\Pi}$ with  $g\in\mathcal{F}_{{p{\cal G}}}$, $Q_{{p{\cal G}}}^{(n)}$ is asympto\-tically non-central chi-square  with non-centrality parameter\vspace{-1mm} 
$$
 \frac{64(\dot{\Pi}(0))^2(\Gamma(d/2)((d+1)\frac{\mu_{d;g}\mu_{d+1;g}}{(\mu_{d-1;g})^2}-d\frac{\mu_{d+2;g}}{\mu_{d-1;g}}))^2}{\pi((d^2-1)\Gamma((d-1)/2))^2d^2 {\gamma_{{{\cal G}}}}}\taub_3'\taub_3,
\vspace{-1mm} $$
  where  
   $\taub_3=\lim_{n\rightarrow\infty}\taub^{(n)}_3$
and\vspace{-1mm} 
  $$\gamma_{{{\cal G}}}:=
  \frac{3}{d(d+2)}\frac{\mu_{d+3;g}}{\mu_{d-1;g}} - 2c_{d}^2(d+1)\frac{\mu_{d;g}\mu_{d+2;g}}{(\mu_{d-1;g})^2} + c_{d}^{2}\frac{(d+1)^2}{d}\frac{\mu_{d;g}^{2}\mu_{d+1;g}}{(\mu_{d-1;g})^3}. 
  \vspace{-1mm}$$
\end{thm}
Theorems~\ref{asympt3} and~\ref{propo} allow for computing the desired ARE values as  squared ratios of   local shifts.
 \begin{thm}
Let $f\in \mathcal{F}_1$; suppose that Assumptions (A1-A3) and~(B) hold, and that the skewing function $\Pi$ is continuously differentiable at $0$ with~$\dot{\Pi}(0)\neq0$. Then,   the ARE of   $\phi^{\ddagger (n)}_f$ with respect to $\phi^{(n)}_{{p{\cal G}}}$ under local alternatives of the form ${\rm P}^{(n)}_{\thetab,\Sigmab,n^{-1/2}\taub_{3}^{(n)};g,\Pi}$ with $ g \in \mathcal{F}_{1;f} \cap \mathcal{F}_{{p{\cal G}}}$ is\vspace{-1mm}  
$$
\text{ARE}_g(\phi^{\ddagger (n)}_f/\phi^{(n)}_{{p{\cal G}}}) =
\displaystyle \frac{d^3\pi \left(1- {\alpha_{d,f,g}}/{{\mathcal{K}_{d,f,g}}}\right)^2\left((d^2-1)\Gamma((d-1)/2)\right)^2 {\gamma_{{{\cal G}}}}}{16\left(\Gamma(d/2)((d+1)\frac{\mu_{d;g}\mu_{d+1;g}}{(\mu_{d-1;g})^2}-d\frac{\mu_{d+2;g}}{\mu_{d-1;g}})\right)^2\gamma_{d,f,g}}.
\vspace{-3mm}$$
\end{thm}

\begin{table}[H]
\small
\centering
  \caption{
  AREs, with respect to  $\phi^{(n)}_{{p{\cal G}}}$ and under several skew-$t$ alternatives, of our tests $\phi^{\ddagger (n)}_{t_{\nu}}$ for various values of $\nu$ and the dimension $d$.
}
{
\footnotesize
\begin{tabular}{@{}ccccccc}\hline
& &\multicolumn{5}{c} {Degrees of freedom of the underlying $t$ density} 
\\[.5mm]
\cmidrule{3-7}
$d$&test&4.1 & 5& 7& 10& 20\vspace{1mm}\\
 2&$\phi^{\ddagger (n)}_{t_{4}}$&10.968&1.964&1.305&1.156&1.085\\
  &$\phi^{\ddagger (n)}_{t_{5}}$&10.912&1.978&1.342&1.208&1.155\\
  &$\phi^{\ddagger (n)}_{t_{7}}$&10.630&1.955&1.358&1.249&1.223\\
    &$\phi^{\ddagger (n)}_{t_{10}}$&10.172&1.892&1.345&1.261&1.264\\
&$\phi^{\ddagger (n)}_{t_{20}}$&8.997&1.705&1.262&1.231&1.287\\
 3&$\phi^{\ddagger (n)}_{t_{4}}$&11.780&2.149&1.473&1.341&1.300\\
  &$\phi^{\ddagger (n)}_{t_{5}}$&11.725&2.164&1.511&1.397&1.383\\
  &$\phi^{\ddagger (n)}_{t_{7}}$&11.449 &2.140&1.528&1.442&1.462\\
  &$\phi^{\ddagger (n)}_{t_{10}}$&10.993&2.076&1.513&1.455&1.510\\
    &$\phi^{\ddagger (n)}_{t_{20}}$&9.804&1.882&1.424&1.420&1.539\\
 5&$\phi^{\ddagger (n)}_{t_{4}}$&12.867&2.410&1.729&1.646&1.706\\
  &$\phi^{\ddagger (n)}_{t_{5}}$&12.818&2.423&1.765&1.703&1.794\\
   &$\phi^{\ddagger (n)}_{t_{7}}$&12.564&2.401&1.783&1.751&1.886\\
  &$\phi^{\ddagger (n)}_{t_{10}}$&12.132&2.338&1.767&1.766&1.945\\
    &$\phi^{\ddagger (n)}_{t_{20}}$&10.964 &2.141&1.670&1.724&1.983\\
 10&$\phi^{\ddagger (n)}_{t_{4}}$&7.486 &2.759&2.117&2.170&2.548\\
  &$\phi^{\ddagger (n)}_{t_{5}}$&14.202&2.770&2.143&2.215&2.626\\
  &$\phi^{\ddagger (n)}_{t_{7}}$&14.008&2.752&2.158&2.256&2.719\\
  &$\phi^{\ddagger (n)}_{t_{10}}$&13.654&2.699&2.143&2.270&2.786\\
    &$\phi^{\ddagger (n)}_{t_{20}}$&12.618&2.519&2.047&2.224&2.832\\
\hline  
\end{tabular}
  }
\label{ares}
\end{table}

Table~\ref{ares} provides numerical values of the AREs for various skew-$t$ alternatives. All ARE values are larger than one, sometimes quite significantly; as a rule, they decrease with the degrees of freedom of the underlying Student, and increase with the dimension. The test for which the reference $f$ coincides with the actual $g$ yields the maximal value of~$ARE_g$, as it should. Note that  we deliberately opted for the test $\phi^{\ddagger (n)}_{t_{4}}$ instead of $\phi^{\ddagger (n)}_{t_{4.1}}$: hence,  the highest values of $ARE_{t_{4.1}}$  are not shown here.

\section{Comparative finite-sample  study}\label{numsec}

In this section we investigate, via Monte Carlo simulations, the finite-sample properties of the tests we are proposing  and some of their competitors---first for specified location (Section~\ref{simuspec}) and then for unspecified location (Section~\ref{simuunspec}). We   start with a brief description of the competing methods to be considered in this study.

\subsection{Competing methods: specified location}\label{simucompet}
Most   tests proposed in the literature are dealing with the specified-location problem
 We selected the following two, proposed by Baringhaus~\cite{baringhaus1991testing} and Cassart \cite{Cassart07}, respectively. 

%


 (a) Baringhaus~\cite{baringhaus1991testing}  proposes a class of  tests $\phi_{\text{Bar},\thetab}^{(n)}$
  based on\vspace{-2mm} 
\begin{equation}\label{Baring}B^{(n)}:=\frac{1}{n^2}\sum_{i,j=1}^nh(\Ub_i(\thetab,\hat\Sigmab)'\Ub_j(\thetab,\hat\Sigmab))(n-\max\left({R_i,R_j}\right)+1),
\vspace{-2mm}\end{equation}
where~$h$ is defined over~$[-1,1]$ and satisfies some regularity conditions,~$\hat\Sigmab$ is Tyler's estimator of scatter, and    $R_i$ is the rank of~$\|\hat\Sigmab^{-1/2}(\Xb_i-\thetab)\|$ among~$\|\hat\Sigmab^{-1/2}(\Xb_1-\thetab)\|,\ldots,\|\hat\Sigmab^{-1/2}(\Xb_n-\thetab)\|$. In our simulations we chose~$h(t)=\big(\frac{2}{17/8-t}\big)^{1/2}-1$,  $t\in[-1,1]$ because the asymptotic null distribution of $B^{(n)}$ then coincides (up to a multiplicative constant) with that of the squared   Kolmogorov-Smirnov statistic for the problem under study (other choices of~$h$ would require simulation-based approximations of limiting null distributions). No moment assumptions are required. Baringhaus~\cite{baringhaus1991testing} actually introduced $\phi_{\text{Bar},\thetab}^{(n)}$ as a test for spherical  symmetry (with ${\bf I}_d$ instead of $\hat\Sigmab$ in~\eqref{Baring}). Empirical sphericization via the Tyler estimator $\hat\Sigmab$ turns it into a test for  elliptical symmetry; this has been proposed by \cite{dyckerhoff2015depth} who establishes (via simulations) the validity of the procedure in dimension~$d=2$.  

(b) The pseudo-Gaussian  tests $\phi_{p{\cal G},\thetab}^{(n)}$ described by  \cite{Cassart07}  achieve Le Cam optimality against the Fechner-type multinormal alternatives defined there (Chapter 3). When the location $\thetab$ is known, the test $\phi_{p{\cal G},\thetab}^{(n)}$ rejects the  hypothesis of elliptical symmetry  with location $\thetab$ at asymptotic level~$\alpha$ whenever \vspace{-1mm}
$$\frac{8}{3nm^{(n)}_4}\sum_{i,j=1}^n  (d_i(\thetab,\hat\Sigmab))^2 (d_j(\thetab,\hat\Sigmab))^2\SSbb'_{{\Ub}_i(\thetab,\hat\Sigmab)}\SSbb_{{\Ub}_j(\thetab,\hat\Sigmab)}
\vspace{-1mm}$$ 
($\SSbb_{{\Ub}_i(\thetab,\hat\Sigmab)}$ defined in \eqref{Sdef}) exceeds the $(1-\alpha)$ chi-quare quantile $\chi^2_{d; 1-\alpha}$.  Finite moments of order four are required. 
 For~$\hat{\pmb\Sigma}$, we still use Tyler's estimator.

\subsection{Competing methods: unspecified location}\label{simucompet}
The list of competitors is  shorter in the unspecified-location case---despite the importance of the problem. Below, we are considering the unspecified-location pseudo-Gaussian tests $\phi_{p{\cal G}}^{(n)}$  proposed by Cassart~\cite{Cassart07}, the Schott test  $\phi_{{\text{Schott}}}^{(n)}$ \cite{schott2002testing}, and the Koltchinskii--Sakhanenko test $\phi_{\text{K-S}}^{(n)}$  \cite{koltchinskii2000testing}. 

(c) Cassart's location-unspecified test $\phi_{p{\cal G}}^{(n)}$ is described in Section~\ref{ARE}, where we refer to for details; its validity requires finite moments of order~four.

(d)  Schott's test $\phi_{{\text{Schott}}}^{(n)}$ \cite{schott2002testing} involves a test statistic based on fourth-order moments;  its validity requires finite eighth-order moments.  The underlying idea is that the fourth-order moment structure of an elliptical distribution is   a scalar multiple of that of   a normal distribution. Therefore, to test whether a given population has an elliptical distribution, it is sufficient to test whether its fourth-order moment structure matches  that of a Gaussian population. A closed-form of the test statistic involves a long list of notations which we are skipping here---see~\cite{schott2002testing} for details; its asymptotic distribution    is chi-square with~$d^2+ {d(d-1)(d^2+7d-6)}/{24}-1$ degrees of freedom. 

(e) The Koltchinskii--Sakhanenko \cite{koltchinskii2000testing} test statistics $\phi_{\text{K-S}}^{(n)}$ are obtained  as functionals of empirical processes indexed by special classes of functions. Let~$\mathcal{F}_B$ be a class of Borel functions from $\Rd$ to $\R$. Their test statistics are functionals (for example, sup-norms) of the stochastic process\vspace{-1mm}
$$
n^{-1/2}\sum_{i=1}^{n}\left(f(\hat\Sigmab^{-1/2}(\Xb_i-\hat\thetab)) - m_{f}(d_i(\hat\thetab,\hat\Sigmab))\right), \vspace{-1mm}
$$
where $f \in \mathcal{F}_B$, $m_f(\rho)$ is the average value of $f$ on the sphere with radius~$\rho>0$,  and $\hat\thetab$ and $\hat\Sigmab$ denote the sample average and covariance matrix,  respectively. Several examples of classes $\mathcal{F}_B$ and test statistics based on the sup-norm of the above process are considered in~\cite{koltchinskii2000testing}. Here we restrict  to~$\mathcal{F}_B:=\left\{I_{0<||\xb||\leq t}\psi\left(\frac{\xb}{||\xb||}\right):\psi\in G_l,||\psi||_2\leq1,t>0\right\}$ where $I_{A}$ stands for the indicator function of   $A$,  $G_l$ for the linear space of spherical harmonics of degree less than or equal to $l$ in $\R^d$, and~$||\cdot||_2$ is the~$L^2$-norm on    the unit sphere $\mathcal{S}^{d-1}$ in $\R^d$. Critical values are obtained via a  bootstrap procedure.

\subsection{Finite-sample performance:  specified location (Table~\ref{TlocSpec1})}\label{simuspec}

Without loss of generality, fix $\thetab=\zerob$. In order to compare the null and non-null finite-sample behavior of our optimal semiparametric test $\phi^{(n)}_{\zerob}$ with that of the Baringhaus and pseudo-Gaussian tests $\phi_{\text{Bar},\thetab}^{(n)}$ and $\phi_{p{\cal G},\thetab}^{(n)}$, we consider samples of size~$n=100$ from various distributions in dimension $d=3$, and calculate their rejection frequencies on the basis of $N=3000$ replications. Under the null hypothesis, we consider the three-dimensional normal   and Student $t$ elliptical distributions with $\nu= $2.1, 4.1, and 8 degrees of freedom, all with scatter~$
\Sigmab = 
\begin{bmatrix}
    2 & 1 & 1\\
   1 & 3 & 2  \\
   1&2&5 
\end{bmatrix}
$; the degrees of freedom~2.1 and~4.1 were selected as having finite moments of orders 2 and 4, respectively. 

Alternatives are of four different types:  normal   and Student skew-elliptical  (increasing $\lambdab$ values) in Table~\ref{TlocSpec1}(a),  sinh-arcsinh- (SAS-) transformed   normal and~$t_{4.1}$ (same $\Sigmab$ matrix as above; skewness parameters as indicated;   kurtosis parameters  all fixed to 1), location-scale  Gaussian mixtures (LSGM), and mixtures of Gaussian distributions in Table~\ref{TlocSpec1}(b). 

The skew-elliptical  alternatives are those against which $\phi^{(n)}_{\zerob}$ is optimal. The sinh-arcsinh-transformed families are families of skewed distributions in dimension $d$ (see \cite{jones2009sinh})   indexed by a $d$-dimensional parameter $\lambdab$ with the same interpretation as in skew-elliptical families.  

As proposed by \cite{wraith2015location}, we are considering a particular case of multivariate location-scale  Gaussian mixtures (LSGM) yielding the so-called {\it multiple scaled generalized hyperbolic} (MSGH) distributions. Those distributions are indexed by parameters $\mub$, $\bf D$, $\Ab$, $\betab$, $\lambdab$, $\gammab$, and $\deltab$. More specifically, in Table~\ref{TlocSpec1}(b),  we chose the three-dimensional  MSGH  with $\gammab = (2,2,2)'$, $\deltab = (1,1,1)'$, $\lambdab = (-1/2,2,1)'$, $\Ab = \Ib_3$, and~$
{\bf D} =
\begin{bmatrix}
    \sqrt{2}/2 & -\sqrt{2}/2 & 0\\
  \sqrt{2}/2 &\sqrt{2}/2 & 0\\
  0&0&1
\end{bmatrix}
$. 
Finally, the  mixtures of Gaussian distributions  in Table~\ref{TlocSpec1}(b) are of the form~$\frac{1}{2}{\cal N}_3\big(\mub_1,\Sigmab_1\big)+\frac{1}{2}{\cal N}_3\big(\mub_2,\Sigmab_2\big)$, with various locations $\mub_1$ and  $\mub_2$ and scatter matrices  
$\Sigmab_1
= 
\begin{bmatrix}
    2 & 1 & 1\vspace{-0.75mm}\\
   1 & 3 & 2 \vspace{-0.75mm} \\
   1&2&5\vspace{-0.75mm} 
\end{bmatrix}
$ 
 and~$
\Sigmab_2 
= 
\begin{bmatrix}
    1 & 0.5 & 0.5\vspace{-0.75mm} \\
   0.5 & 1 & 0.5 \vspace{-0.75mm}  \\
   0.5&0.5&1 \vspace{-0.75mm} 
\end{bmatrix}
$, respectively.  
For each case, we considered increasingly skewed alternatives.

\begin{table}
\small
\centering
  \caption{
 Rejection frequencies (out of $N=3,000$ replications), under {\rm (a)} various three-dimensional elliptical (${\lambdab = (0,0,0)}$) and related skewed densities (increasing $\lambdab$ values) and {\rm (b)} skewed  SAS-normal, SAS-$t_{4.1}$, location-scale Gaussian mixtures (LSGM) and location Gaussian mixtures (increasing $\lambdab$) values), of our optimal specified-location($\thetab_0=\zerob$)  test $\phi^{(n)}_{\zerob}$, the Baringhaus test~$\phi_{\text{Bar},{\zerob}}^{(n)}$, and Cassart's   pseudo-Gaussian test $\phi_{p{\cal G},\zerob}^{(n)}$. The sample size is~$n=100$, the nominal probability level~5\%.
}
{  
\footnotesize
\begin{tabular}{@{}rcccccc}\toprule 
 (a)\ \    $\lambdab^\prime$ &     $(0,0,0)$    & \hspace{4mm}   $(0.1,-0.2,0)$       & \hspace{5mm}  $(0.3,-0.6,0)$   & \hspace{0.002mm}     $(0.1,0.1,0.1)$     &   \hspace{-3mm}    $(0.2,0.2,0.2)$  & \hspace{0.5mm}  $(0.3,0.3,0.3)$       \\ 
  \cmidrule(lr){2-7}
{\footnotesize test}& \multicolumn{6}{c} {\footnotesize Skew-normal}   \\[.5mm]
 \cmidrule(lr){2-7}
$\phi^{(n)}_{\zerob}$ &0.055&0.193&0.934&0.293&0.847& 0.992  \\
$\phi_{\text{Bar},{\zerob}}^{(n)}$  &0.038&0.088&0.625& 0.132&0.467& 0.831\\
$\phi_{p{\cal G},\zerob}^{(n)}$   &0.055&0.165&0.873&0.243& 0.756& 0.975	   \\
 \multicolumn{7}{c} {\footnotesize Skew-$t_{2.1}$}   \\[.5mm]
  \cmidrule(lr){2-7}
$\phi^{(n)}_{\zerob}$  &0.039& 0.106 &  0.651& 0.145& 0.522& 0.815\\
$\phi_{\text{Bar},{\zerob}}^{(n)}$ & 0.035&0.092 &0.675 & 0.147& 0.521& 0.874  \\
$\phi_{p{\cal G},\zerob}^{(n)}$   &0.012&  0.036 &0.157 & 0.040& 0.126& 0.235\\ 
 \multicolumn{7}{c} {\footnotesize Skew-$t_{4.1}$}   \\[.5mm]
  \cmidrule(lr){2-7}
$\phi^{(n)}_{\zerob}$   & 0.040& 0.142&0.864 & 0.239& 0.753& 0.964\\
$\phi_{\text{Bar},{\zerob}}^{(n)}$  &0.037& 0.090& 0.662& 0.131& 0.501& 0.857 	  \\
$\phi_{p{\cal G},\zerob}^{(n)}$  & 0.034& 0.078&0.460 & 0.121& 0.377& 0.650\\ 	
 \multicolumn{7}{c} {\footnotesize Skew-$t_{8}$}   \\[.5mm]
  \cmidrule(lr){2-7}
$\phi^{(n)}_{\zerob}$  &0.050&0.170 &0.902 & 0.277&  0.813& 0.990  \\
$\phi_{\text{Bar},{\zerob}}^{(n)}$ &0.034& 0.081& 0.638& 0.131& 0.475& 0.862	  \\
$\phi_{p{\cal G},\zerob}^{(n)}$ &0.042& 0.119& 0.688& 0.186& 0.587& 0.879	   \\
\bottomrule
\end{tabular}

\centering 
\begin{tabular}{@{}ccccccc}\toprule
  (b)\ \    $\lambdab^\prime$ & 
   $ (0,0,0)$  &    
   $ (0.05,-0.1,0.05)$&
   $ (0.15,-0.3,0.15)$ &    
   $(0.1,0.1,0.1)$     &     
   $ (0.2,0.2,0.2)$  & 
   ${(0.3,0.3,0.3)}$    \\
 \cmidrule(lr){2-7}
{\footnotesize test}& \multicolumn{6}{c} {\footnotesize SAS-normal}   \\[.5mm]
  \cmidrule(lr){2-7}
$\phi^{(n)}_{\zerob}$  &0.057& 0.276&0.997 & 0.170& 0.587& 0.935   \\
$\phi_{\text{Bar},{\zerob}}^{(n)}$ &0.034& 0.065&0.553 & 0.038& 0.0583& 0.113	  \\
$\phi_{p{\cal G},\zerob}^{(n)}$   &0.052&0.363 &0.999 & 0.273& 0.842& 0.998\\
 \multicolumn{7}{c} {\footnotesize SAS-$t_{4.1}$}   \\[.5mm]
  \cmidrule(lr){2-7}
$\phi^{(n)}_{\zerob}$  &0.044&0.257 &0.992 & 0.166& 0.548& 0.893   \\
$\phi_{\text{Bar},{\zerob}}^{(n)}$  &0.029& 0.133& 0.874& 0.058& 0.119& 0.290 	  \\
$\phi_{p{\cal G},\zerob}^{(n)}$   &0.038& 0.159&0.834 & 0.117& 0.441& 0.752		   \\
 \multicolumn{7}{c} {\footnotesize LSGM}   \\[.5mm]
  \cmidrule(lr){2-7}
$\phi^{(n)}_{\zerob}$  &0.048& 0.098& 0.539& 0.176& 0.601& 0.926  \\
$\phi_{\text{Bar},{\zerob}}^{(n)}$  &0.034&0.063 &0.409 & 0.117& 0.460& 0.851	  \\
$\phi_{p{\cal G},\zerob}^{(n)}$   &0.121& 0.166&0.368 & 0.210&  0.418& 0.677	   \\
 \multicolumn{7}{c} {\footnotesize Gaussian Mixture}   \\[.5mm]
 \cmidrule(lr){2-7} \multicolumn{1}{r}{$\mub_1^\prime$} &$(0,0,0)$ & $(0,0,0)$ &$(0,0,0)$& $(1,0,0)$ & $ (1,0,0)$& $ (1,0,0)$\\
 
\multicolumn{1}{r}{$\mub_2^\prime$}&$(0,0,0)$ &$ (-0.5,0,0)$& $ (-1,0,0)$& $(-1,0,0)$ & $ (-2,0,0)$ & $(-3,0,0)$\\
 \cmidrule(lr){2-7}
$\phi^{(n)}_{\zerob}$  &0.048& 0.379 & 0.936&0.050&0.474& 0.939	   \\
$\phi_{\text{Bar},{\zerob}}^{(n)}$  &0.034& 0.080&0.264&0.710&0.395&0.340	  \\
$\phi_{p{\cal G},\zerob}^{(n)}$   &0.052&0.272 & 0.836&0.063& 0.417&0.893 \\
 \bottomrule
  \end{tabular}
  }
\label{TlocSpec1}
\end{table}

Inspection of Table ~\ref{TlocSpec1} indicates   that $\phi^{(n)}_{\thetab}$ (here~$\phi^{(n)}_{\bf 0}$) uniformly satisfies\footnote{Within the confidence limits of the Monte Carlo experiment: with 3000 replications, a 5\% confidence interval   centered at the   rejection frequencies shown in all tables in this section has approximate  length 0.015.} the~$5\%$ level constraint and yields excellent powers for almost all settings. It is  outperformed in two cases only: 

{\it (i)} by Baringhaus'~$\phi_{\text{Bar},{\zerob}}^{(n)}$ test under skew-elliptical $t_{2.1}$; the same $\phi_{\text{Bar},{\zerob}}^{(n)}$, however, is much weaker under all other skew distributions; this might be due to slow convergence,  under heavy tails, to  limit distributions;

\noindent  \ \  {\it (ii)} by  Cassart's pseudo-Gaussian test~$\phi_{p{\cal G},\zerob}^{(n)}$ under  SAS-normal and LSGM distributions---the latter case, however, is explained by severe over-rejection (rejection frequency 21\% at 5\% nominal level!) under the null.

The results under Gaussian mixtures (bottom of Table~\ref{TlocSpec1}) deserve some further comments. Note that the corresponding first column does not address a null hypothesis situation: although $\mub_1=\zerob=\mub_2$, the resulting mixture is not elliptical. A comparison between columns~3 ($\mub_1=(0,0,0)^\prime$, $\mub_2=(-1,0,0)^\prime$) and~4 ($\mub_1=(1,0,0)^\prime$, $\mub_2=(-1,0,0)^\prime$)  is particularly intriguing. The distribution in column~4 indeed is strictly  ``less elliptical'' than in column~3; nevertheless, the power of $\phi^{(n)}_{\bf 0}$, which is almost one in column~3, reduces to the nominal level in column 4.  This is an illustration of the fact that specified-location  tests cannot be considered as genuine ellipticity tests (see Section~\ref{pitsec}).  Baringhaus apparently is less sensitive to that   phenomenon---at the price, however, of very low powers under most values of $\mub_1\neq\mub_2$.

\subsection{Finite-sample performance: unspecified location (Tables~\ref{TnotSpecdim2_1})   and~\ref{TnotSpecdim3_1}--\ref{TnotSpecdim3_2})}\label{simuunspec}
The tests considered here are our optimal tests $\phi^{\ddagger (n)}_{f}$  ($f$   elliptical Student  with $\nu=2.1$, 4,  and  8 degrees of freedom), Schott's  test~$\phi_{\text{Schott}}^{(n)}$, Cassart's pseudo-Gaussian test~$\phi_{p{\cal G}}^{(n)}$, and  Koltchinskii and Sakha\-nenko's~$\phi_{\text{K-S}}^{(n)}$ test. Table~\ref{TnotSpecdim2_1} is dealing with dimension $d=2$,  Tables~\ref{TnotSpecdim3_1} and~\ref{TnotSpecdim3_2} (in Appendix~D) with $d=3$.  Because of  its computational complexity, the Koltchinskii-Sakha\-nenko 
 test is considered  for $d=2$~only. 

We still consider samples of size $n=100$, from the same distributions\footnote{
 For $d = 2$, we considered the MSGH distribution   with   parameter values~$\pmb{\gamma} = (2,2)'$, $\pmb{\delta} = (1,1)'$, $\lambdab = (-1/2,2)'$, ${\bf A} =\Ib_2$, and 
${\bf D} =
 \begin{bmatrix}
   \sqrt{2+\sqrt{2}}/2 & - \sqrt{2-\sqrt{2}}/2\\
 \sqrt{2-\sqrt{2}}/2 &  \sqrt{2+\sqrt{2}}/2
\end{bmatrix}$.} 
  as   in~\ref{simuspec}, 
and calculate the rejection frequencies on the basis of $N=3000$ replications. 
Again, our tests $\phi^{\ddagger (n)}_{f}$  outperform the other tests for almost all settings. 
 The pseudo-Gaussian test performs very well for the SAS-normal distribution. In all other settings, the $\phi^{\ddagger (n)}_{t_\nu}$  tests yield the best results. Quite remarkably,   $\phi^{\ddagger (n)}_{f}$ under Gaussian mixtures does not suffer at all  the problems its specified-location counterpart was exhibiting in Table~\ref{TlocSpec1}, and uniformly dominates all its  competitors.

 \subsection{The pitfalls of specified-location tests}\label{pitsec} We already stressed the fact that most tests available in the literature are dealing with the null hypothesis of specified-location ellipticity. Those tests, as a rule, are reasonably powerful at detecting either elliptical location alternatives (a simple shift in the null distribution)
  or  fixed-location violations of ellipticity. Problems occur when both violations are present, with opposite impacts on the test statistic:  powers then completely collapse. 
 
  To showcase this, we ran   our tests $\phi^{(n)}_{\zerob}$ and $\phi_{t_4}^{\ddagger;(n)}$ against simulated (3000 replications) 10-dimensional\footnote{The higher the dimension, the more serious the problem.} Gaussian mixtures of the form
$$ 0.8\,{\cal N}_{10}(10\, \eb_{1;10}, {\bf I}_{10}) + 0.2\,{\cal N}_{10}(-10\, \eb_{1;10}, {\bf I}_{10})-\deltab_i,\qquad  i=1,2,
$$

\begin{landscape}

\begin{table}
\small\vspace{-12mm}
\centering
  \caption{
  Rejection frequencies (out of $N=3,000$ replications), under various bivariate elliptical (${\lambdab = (0,0)}$) and related skewed densities (increasing $\lambdab$ values), of our unspecified-location optimal tests~$\phi_f^{\ddagger;(n)}$ ($f$ the bivariate $t$ distributions with~2.1,~4, and~8 degrees of freedom), Schott 's test $\phi_{\text{Schott}}^{(n)}$, Cassart's pseudo-Gaussian test~$\phi_{p{\cal G}}^{(n)}$, and the Koltchinskii--Sakhanenko test~$\phi_{\text{K-S}}^{(n)}$ for the null hypothesis of ellipticity with unspecified location. The sample size is~$n=100$, the nominal probability level~5\%.
}
{
\footnotesize
\begin{tabular}{@{}lcccccccccc}
\toprule
(b)\ \  $\lambdab^\prime$  &     $(0,0)$ &$(1,-1)$   &    $(1,1)$       & $(2,2)$ &$(3,3)$ &     $(0,0)$ &$(0.15,-0.2)$   &    $(0.15,0.15)$       & $(0.3,0.3)$ &$(0.45,0.45)$    \\
\cmidrule(lr){2-6}\cmidrule(lr){7-11}
test& \multicolumn{5}{c} {\footnotesize Skew-normal} &\multicolumn{5}{c} {\footnotesize SAS-normal}   \\[.5mm]
 \cmidrule(lr){2-6}\cmidrule(lr){7-11}
$\phi^{\ddagger (n)}_{t_{2.1}}$  &0.044& 0.052&0.129& 0.501& 0.684&0.049& 0.254& 0.229& 0.732& 0.973 \\ 
$\phi^{\ddagger (n)}_{t_{4}}$ &0.043&0.053 & 0.131& 0.510& 0.691&0.046&0.261 & 0.233& 0.731& 0.974 \\ 
$\phi^{\ddagger (n)}_{t_{8}}$   &0.044& 0.055& 0.129& 0.502& 0.679& 0.049& 0.255& 0.230& 0.710& 0.964\\
$\phi_{\text{Schott}}^{(n)}$  & 0.034 &0.040 &  0.035     &0.036 &  0.042 &0.040 & 0.035 &  0.034   &0.028 &  0.034	  \\
$\phi_{p{\cal G}}^{(n)}$  &0.045& 0.040& 0.064& 0.139& 0.192 &0.051&0.420 & 0.284& 0.808& 0.985 \\
$\phi_{\text{K-S}}^{(n)}$ &0.048&0.047&0.065&0.095&0.116&0.056&0.096&0.069&0.121&0.213\\
& \multicolumn{5}{c} {\footnotesize Skew-$t_{2.1}$}& \multicolumn{5}{c} {\footnotesize SAS-$t_{4.1}$}    \\[.5mm]
\cmidrule(lr){2-6}\cmidrule(lr){7-11}   
$\phi^{\ddagger (n)}_{t_{2.1}}$  &0.033&0.294 & 0.445& 0.621& 0.675&0.040& 0.168& 0.120& 0.368& 0.698\\
$\phi^{\ddagger (n)}_{t_{4}}$   &0.030& 0.230& 0.349& 0.509& 0.561&0.038 & 0.139& 0.109& 0.325& 0.618\\
$\phi^{\ddagger (n)}_{t_{8}}$   &0.022&0.165 & 0.260&0.376& 0.431 &0.037&0.119 & 0.088& 0.267& 0.511   \\
$\phi_{\text{Schott}}^{(n)}$    & 0.265 & 0.285   & 0.309  & 0.340 & 0.324& 0.060 &0.061  & 0.059  & 0.061& 0.067 \\
$\phi_{p{\cal G}}^{(n)}$  &0.020& 0.101& 0.152& 0.221& 0.265	 &0.037& 0.171& 0.096& 0.308& 0.578 \\
$\phi_{\text{K-S}}^{(n)}$  &0.057&0.211&0.341&0.473& 0.539	    &0.059&0.086&0.075&0.131&0.229 \\
& \multicolumn{5}{c} {\footnotesize Skew-$t_{4.1}$}& \multicolumn{5}{c} {\footnotesize LSGM}    \\[.5mm]
\cmidrule(lr){2-6}\cmidrule(lr){7-11} 
$\phi^{\ddagger (n)}_{t_{2.1}}$     &0.043&0.291 & 0.535& 0.846& 0.912&0.038& 0.085&  0.094& 0.188& 0.377 \\
$\phi^{\ddagger (n)}_{t_{4}}$    &0.040&0.266 & 0.482& 0.775& 0.844&0.038& 0.076& 0.079& 0.162& 0.328 \\
$\phi^{\ddagger (n)}_{t_{8}}$ & 0.036& 0.222& 0.409& 0.675& 0.734&0.036& 0.052& 0.072& 0.135& 0.268\\
$\phi_{\text{Schott}}^{(n)}$  &  0.058 & 0.064  & 0.067  & 0.085& 0.093 & 0.285 & 0.288& 0.289 & 0.278&  0.290 \\
$\phi_{p{\cal G}}^{(n)}$   & 0.035&0.153 & 0.244& 0.369& 0.386 &0.033&0.058 & 0.053& 0.108& 0.209 \\
$\phi_{\text{K-S}}^{(n)}$  &0.056& 0.082&0.110&0.179&0.213&0.241& 0.251&0.245&0.293&0.334\\
& \multicolumn{5}{c} {\footnotesize\phantom{ Skew-$t_{8}$}} & \multicolumn{5}{c} {\footnotesize Gaussian Mixture}    \\[.5mm]
\cmidrule(lr){7-11}
& \multicolumn{5}{c} {\footnotesize Skew-$t_{8}$}&$\mub_1^\prime$ $ (0,0)$ & $  (0,0)$& $(1,0)$ & $(1,0)$& $(1,0)$\\
 &&&&&&$\mub_2^\prime$ $(0,0)$ & $ (-1,0)$& $(-1,0)$ & $(-2,0)$ & $ (-3,0)$\\
 \cmidrule(lr){2-6} \cmidrule(lr){7-11}
$\phi^{\ddagger (n)}_{t_{2.1}}$   & 0.046&0.170 & 0.374& 0.767& 0.871&0.044& 0.199& 0.475&0.575&0.552\\
$\phi^{\ddagger (n)}_{t_{4}}$ & 0.046& 0.167 & 0.357& 0.734& 0.845 &0.043&0.199&0.452&0.520&0.482\\
$\phi^{\ddagger (n)}_{t_{8}}$  &0.045& 0.158& 0.326& 0.674& 0.790&0.046&0.192&0.407&0.444&0.391 \\
$\phi_{\text{Schott}}^{(n)}$ & 0.043 & 0.030& 0.033& 0.036&    0.054 & 0.082 & 0.085& 0.126 & 0.276&  0.570  \\
$\phi_{p{\cal G}}^{(n)}$ &0.038& 0.101&  0.175& 0.312& 0.365	  & 0.036&0.102&0.151&0.123&0.092  \\
$\phi_{\text{K-S}}^{(n)}$  &0.052&0.053&0.080&0.117&0.146&0.049&0.077&0.110&0.162&0.306   \\
 \bottomrule 
  \end{tabular}
  }
\label{TnotSpecdim2_1}
\end{table}
\end{landscape}

\noindent with $\eb_{1;10}=(1,0,0,0,0,0,0,0,0,0)'$, $\deltab_1=(-6,0,0,0,0,0,0,0,0,0)'$, and~$\deltab_2=(6,0,0,0,0,0,0,0,0,0)'$, that is, we consider two distinct scenarios, {\it (a)} and {\it (b)}, say. 
Note that non-ellipticity is strictly the same under both scenarios: only locations differ, with {\it (b)} remaining centered at $\zerob$. The rejection frequencies are as follows: under {\it (a)}, $\phi^{(n)}_{\zerob}$  reaches 0.042 and $\phi_{t_4}^{\ddagger;(n)}$~0.681, while under {\it (b)},  $\phi^{(n)}_{\zerob}$ reaches 1.000 and $\phi_{t_4}^{\ddagger;(n)}$ 0.685. It appears very clearly that the unspecified-location test makes no distinction between {\it (a)} and {\it (b)},  detecting asymmetry under both, while the location-specified test fails to detect    non-ellipticity under {\it (a)}. The reason is that non-ellipticity and   location shift under {\it (a)} have opposite effects on the test statistic, which cancel each other. On the contrary, under {\it (b)}, the specified-location test is stronger as it does not suffer from the loss of power due to the estimation of $\thetab$.

The conclusion is that one should  be extremely cautious before concluding that ellipticity can or cannot be rejected on the basis of a specified-location test, and rather check whether the unspecified-location procedure does not lead to the opposite conclusion. This warning is all the more important in higher dimensions, where a plot of the observations does not help much: Section~\ref{empirsec} provides a real-life example of this in dimension $d=17$.

\section{An empirical analysis of financial returns data}\label{empirsec}
 Elliptical symmetry with respect to the origin  is a common assumption in the multivariate analysis of financial data.  In this section, we are  testing whether such assumption is acceptable on a dataset   consisting of 18 years of daily returns from 17 major   financial indexes from  America (S$\&$P500, NASDAQ, TSX, Merval, Bovespa and IPC), Europe/Middle East (AEX, ATX, BEL, DAX and CAC40), and East Asia/Oceania (HgSg, Nikkei, 
BSE, KOSPI , TSEC and AllOrd). The sample consists of 4619 observations, from January 7, 2000 through 
September~20,~2017. Those observations, of course, are serially dependent. In order to  neutralize conditional  heteroskedasticity,  following the suggestion of \cite{lombardi2009indirect} for elliptical and possibly heavy-tailed data, they were adjusted via  AR(2)-GARCH(1,1)  filtering.

We shall test for elliptical symmetry both about the fixed location $\thetab_0=\zerob$ (a natural choice) and without specifying the center of symmetry. We thus  compare our test $\phi^{(n)}_{\zerob}$ with our test $\phi^{\ddagger (n)}_{t_4}$ based on the elliptical~$t$ distribution with 4 degrees of freedom. For the entire 17-dimensional data set, we obtain for $\phi^{(n)}_{\zerob}$ a p-value of 0.18, hence do not reject elliptical symmetry with respect to $\zerob$. If the location is not specified, $\phi^{\ddagger (n)}_{t_4}$, with  p-value virtually  zero, very significantly rejects ellipticity. Now, we  investigate this in more details,  using a rolling window of three years. Table~\ref{datatab} in Appendix~D contains the  p-values corresponding to the resulting 16 three-year periods.  We still observe quite opposite conclusions of the two tests: the specified-location test essentially never rejects,  while the unspecified-location test consistently does. The only explanation for this, which illustrates our warnings from Section~\ref{pitsec}, is that   the actual location is not $\zerob$. The unspecified-location test, in case $\phi^{\ddagger (n)}_{t_4}$ and $\phi^{(n)}_{\zerob}$ yield strongly opposite conclusions,   is thus far more reliable than the specified-location one, from which we can conclude that the assumption of ellipticity in this dataset is unlikely to be satisfied. 

\section{Conclusion}\label{conclu}

Based on a family of generalized skew-elliptical distributions, we are proposing tests for the null hypothesis of elliptical symmetry   under specified and unspecified location, respectively. Theoretical ARE values and finite-sample simulations demonstrate their excellent performance, well beyond the context of skew-elliptical alternatives. The inherent  unreliability of  specified-location  methods is stressed.

\bibliography{bibi-1}

\appendix


\newpage 
  
\section*{Appendix A: Proof of Theorem~\ref{ULAN}}

Our proof of  Theorem~\ref{ULAN} relies on  \citet{Swensen85}, Lemma 1---more precisely, on its extension by \citet{GH95}. Checking most of the conditions from \citet{GH95} is  a routine task, which we leave to the reader, and the only difficulty consists in establishing the  quadratic mean differentiability of  $(\thetab,\Sigmab,\lambdab)\mapsto \underline{f}^{1/2}(\xb;\thetab,\Sigmab,\lambdab,f,\Pi)$, which follows from the fillowing lemma.   
\begin{lem}
Letting $f\in \mathcal{F}_1$, suppose that Assumptions (A1) and~(A2) hold and that the skewing function $\Pi$ is continuously differentiable at $0$, with~$\dot{\Pi}(0)\neq0$. Let
\begin{align*}
&D_{\thetab}\underline{f}^{1/2}(\xb;\thetab,\Sigmab,\pmb{0},f,\Pi):=\frac{1}{2}\underline{f}^{1/2}(\xb;\thetab,\Sigmab,f)\,\varphi_f(\|\Sigmab^{-1/2}(\xb-\thetab)\|)\Sigmab^{-1/2}\Ub(\thetab,\Sigmab),\\
&D_{\Sigmab}\underline{f}^{1/2}(\xb;\thetab,\Sigmab,\pmb{0},f,\Pi):=\frac{1}{4}\underline{f}^{1/2}(\xb;\thetab,\Sigmab,f)\Pd(\Sigmab^{\otimes2})^{-1/2}\\
&\quad\qquad\qquad\times\emph{\vect}\left(\psi_f(\|\Sigmab^{-1/2}(\xb-\thetab)\|)\|\Sigmab^{-1/2}(\xb-\thetab)\|\Ub(\thetab,\Sigmab)\Ub'(\thetab,\Sigmab)-\Ik \right),\\
\textrm{and}&\\
&D_{\lambdab}\underline{f}^{1/2}(\xb;\thetab,\Sigmab,\lambdab,f,\Pi)\Big|_{\lambdab=\pmb{0}}:=\underline{f}^{1/2}(\xb;\thetab,\Sigmab,f)\dot{\Pi}(0)\|\Sigmab^{-1/2}(\xb-\thetab)\|\Ub(\thetab,\Sigmab),
\end{align*}
where $\Ub(\thetab,\Sigmab):=\Sigmab^{-1/2}(\xb-\thetab)/\|\Sigmab^{-1/2}(\xb-\thetab)\|$. Then, \vspace{4mm}\small

\noindent {\it (i)} $\!\displaystyle\int\nolimits_{\Rd} \left\{\underline{f}^{1/2}(\xb;\thetab,\Sigmab,{\boldsymbol\ell},f,\Pi)-\underline{f}^{1/2}(\xb;\thetab,\Sigmab,f)-{\boldsymbol\ell}'D_{\lambdab}\underline{f}^{1/2}(\xb;\thetab,\Sigmab,\lambdab,f,\Pi)\Big|_{\lambdab=\pmb{0}}\right\}^2\!d\xb
=~\!o(\|{\boldsymbol\ell}\|^2)\vspace{4mm}$
  \normalsize and \footnotesize
 \begin{align*}
(ii)&\displaystyle \bigints\nolimits_{\Rd} \!\left\{
\vphantom{\begin{pmatrix} D_{\thetab}\underline{f}^{1/2}(\xb;\thetab,\Sigmab,\pmb{0},f,\Pi) \\ D_{\Sigmab}\underline{f}^{1/2}(\xb;\thetab,\Sigmab,\pmb{0},f,\Pi) \\ D_{\lambdab}\underline{f}^{1/2}(\xb;\thetab,\Sigmab,\lambdab,f,\Pi)\Big|_{\lambdab=\pmb{0}} \end{pmatrix}}
\underline{f}^{1/2}(\xb;\thetab+\tb,\Sigmab+\Hb,{\boldsymbol\ell},f,\Pi)-\underline{f}^{1/2}(\xb;\thetab,\Sigmab,f) \right. \\
&\hspace{20mm}
-\left. \begin{pmatrix} \tb \\ \emph{vech}\Hb \\ {\boldsymbol\ell} \end{pmatrix}^\prime\!\begin{pmatrix} D_{\thetab}\underline{f}^{1/2}(\xb;\thetab,\Sigmab,\pmb{0},f,\Pi) \\ D_{\Sigmab}\underline{f}^{1/2}(\xb;\thetab,\Sigmab,\pmb{0},f,\Pi) \\ D_{\lambdab}\underline{f}^{1/2}(\xb;\thetab,\Sigmab,\lambdab,f,\Pi)\Big|_{\lambdab=\pmb{0}} \end{pmatrix}\!  \right\}^2\!\! d\xb 
=o\left(\left\|\begin{pmatrix} \tb \\ \emph{vech}\Hb \\ {\boldsymbol\ell} \end{pmatrix}\right\|^2\right),
\end{align*}
\normalsize
where $\tb\in\Rd$, $\Hb\in\mathcal{S}_d$,   ${\boldsymbol\ell}\in\Rd$, and  $o(\|\cdot\|)$'s are taken for $\|\cdot\|\to0$.
\label{Dlambda}
\end{lem}

\textbf{Proof.}
All $o(\|\cdot\|)$'s below are to be understood as $\|\cdot\|\to0$. Starting with~(i) and letting $\yb:=\Sigmab^{-1/2}(\xb-\thetab)$, the integral takes the form 
\begin{align*}
\nonumber&\int_{\Rd} \left\{\underline{f}^{1/2}(\xb;\thetab,\Sigmab,{\boldsymbol\ell},f,\Pi)-\underline{f}^{1/2}(\xb;\thetab,\Sigmab,f)-{\boldsymbol\ell}'D_{\lambdab}\underline{f}^{1/2}(\xb;\thetab,\Sigmab,\lambdab,f,\Pi)\Big|_{\lambdab=\pmb{0}}\right\}^2\!\! d\xb\vspace{2mm}\\
\nonumber &\quad =\int_{\Rd} \left\{\Pi^{1/2}({\boldsymbol\ell}'\Sigmab^{-1/2}(\xb-\thetab))-\Pi^{1/2}(0)-\Pi^{1/2}(0)\dot{\Pi}(0){\boldsymbol\ell}'\Sigmab^{-1/2}(\xb-\thetab)\right\}^2 \\
\nonumber&\hspace{70mm} \times 2c_{d,f}|\Sigmab|^{-1/2}f(\|\Sigmab^{-1/2}\xb-\thetab\|)d\xb\\
&\quad =\int_{\Rd} \left\{\Pi^{1/2}({\boldsymbol\ell}'\yb)-\Pi^{1/2}(0)-\Pi^{1/2}(0)\dot{\Pi}(0){\boldsymbol\ell}'\yb\right\}^22c_{d,f}f(\|\yb\|)d\yb .
\end{align*}
Since $\Pi(0)=1/2$,  $\Pi^{1/2}(0)\dot{\Pi}(0)=\dot{(\Pi^{1/2})}(0)$. Using the fact that $\Pi$ is bounded, we obtain,  for some real constant $C$, 
\begin{align*}\left\{\Pi^{1/2}({\boldsymbol\ell}'\yb)-\Pi^{1/2}(0)-\Pi^{1/2}(0)\dot{\Pi}(0){\boldsymbol\ell}'\yb\right\}^2 &\leq (2C+ C\Pi(0)(\dot{\Pi}(0))^2\|\yb\|^2)\\
&=: C^+(\|\yb\|^2),\text{ say,}
\end{align*}
where  $\int C^+(\|\yb\|^2) 2c_{d,f}f(\|\yb\|)d\yb <\infty$ since~$f\in\mathcal{F}_1$. 
The result follows from Lebesgue's dominated convergence theorem combined with the fact that
\begin{displaymath}
\left\{\Pi^{1/2}({\boldsymbol\ell}'\yb)-\Pi^{1/2}(0)-\Pi^{1/2}(0)\dot{\Pi}(0){\boldsymbol\ell}'\yb\right\}^2=o(\|{\boldsymbol\ell}\|^2).
\end{displaymath}

Turning to (ii), the integral there is bounded by $C_3(S_1+S_2+\|{\boldsymbol\ell}\|^2S_3)$, where
\begin{align*}
S_1:=&\int_{\Rd} \bigg\{\underline{f}^{1/2}(\xb;\thetab+\tb,\Sigmab+\Hb,f)-\underline{f}^{1/2}(\xb;\thetab,\Sigmab,f) \\ 
&\hspace{40mm}-\begin{pmatrix} \tb \\ \vech\Hb \end{pmatrix}'\begin{pmatrix} D_{\thetab}\underline{f}^{1/2}(\xb;\thetab,\Sigmab,\pmb{0},f,\Pi) \\ D_{\Sigmab}\underline{f}^{1/2}(\xb;\thetab,\Sigmab,\pmb{0},f,\Pi) \end{pmatrix} \bigg\}^2d\xb,\\
S_2:=&\int_{\Rd} \Big\{\underline{f}^{1/2}(\xb;\thetab+\tb,\Sigmab+\Hb,{\boldsymbol\ell},f,\Pi)-\underline{f}^{1/2}(\xb;\thetab+\tb,\Sigmab+\Hb,f)\\
&\hspace{40mm}-{\boldsymbol\ell}'D_{\lambdab}\underline{f}^{1/2}(\xb;\thetab+\tb,\Sigmab+\Hb,\lambdab,f,\Pi)\Big|_{\lambdab=\pmb{0}}\Big\}^2d\xb,\\
S_3:=&\int_{\Rd} \left\|D_{\lambdab}\underline{f}^{1/2}(\xb;\thetab+\tb,\Sigmab+\Hb,\lambdab,f,\Pi)\Big|_{\lambdab=\pmb{0}} \right.\\ 
&\hspace{40mm}\left.-D_{\lambdab}\underline{f}^{1/2}(\xb;\thetab,\Sigmab,\lambdab,f,\Pi)\Big|_{\lambdab=\pmb{0}}\right\|^2d\xb
\end{align*}
and $C_3$ is a strictly positive real constant.

 From Lemma A.1 of \citet{hallin2006semiparametrically1}, we know that, under Assumptions~(A1) and (A2), $S_1$ is  $o\left(\left\|  \tb^\prime , \vech^\prime\Hb \right\|^2\right)$, hence  also~$o\left(\left\|  \tb^\prime ,\vech^\prime\Hb , {\boldsymbol\ell}^\prime  \right\|^2\right)$.  It follows  from  (i) above that the same holds true for $S_2$.  
  It thus remains to show  that $S_3$ is  $o(1)$ to  complete the proof. This, however,  follows from the quadratic mean continuity of~$(\thetab, \Sigmab)\mapsto D_{\lambdab}\underline{f}^{1/2}(\xb;\thetab,\Sigmab,\lambdab,f,\Pi)$, since $||D_{\lambdab}\underline{f}^{1/2}(\xb;\thetab,\Sigmab,\lambdab,f,\Pi)||$ belongs to $L^2(\Rd, d\xb)$ in view of the fact that $f$ admits finite moments of order 2. $\hfill\square$
 
\

\section*{Appendix B: Proof of Theorem~3.1}
   
The following notation will be convenient here and in Appendix C. 
 Letting~$\thetab_n:=\thetab+n^{-1/2}\taub_1^{(n)}$ for some bounded sequence of $d$-dimensional vectors~$\taub_1^{(n)}$ and $\Sigmab_n:=\Sigmab+n^{-1/2}\taub_2^{(n)}$ for some bounded sequence of $d\times d$ matrices~$\taub_2^{(n)}$, define $\varthetab_{0n}:=(\thetab_n^\prime, {\rm vech}^\prime\Sigmab_n,\zerob^\prime)^\prime$. 

   \begin{lem}\label{glemma}
   Let $h:\R^+ \rightarrow \R^+$ be   such that $h^p$ for $p=1$ (resp.,~$p=~\!2$) is integrable with respect to the measure $\nu$,   where $\nu$ is  absolutely continuous with respect to the Lebesgue measure. Then, 
$$
\lim_{n\rightarrow\infty}\int_{\R^d} |h(|| \Sigmab_n^{-1/2}(\xb-\thetab_n)||)-h(|| \Sigmab^{-1/2}(\xb-\thetab)||)|^p d\nu(\xb) =0
$$
for $p=1$  (resp., $p=2$).  

   \end{lem}
   \textbf{Proof.}
For any $\epsilon > 0$, we can choose $h_{\epsilon}$ from $C_{c}^{\infty}(\R^+)$ such\linebreak  that~$||h - h_{\epsilon}||_{L^{p}(d\nu)}<\epsilon$. Then, 
   \begin{align*}
     &\lim_{n\rightarrow\infty} \int_{\R^d} \left|h(|| \Sigmab_n^{-1/2}(\xb-\thetab_n)||)-h(|| \Sigmab^{-1/2}(\xb-\thetab)||)\right|^pd\nu(\xb)\\
     &\leq \lim_{n\rightarrow\infty} \int_{\R^d}\left|h(|| \Sigmab_n^{-1/2}(\xb-\thetab_n)||) - h_{\epsilon}(|| \Sigmab_n^{-1/2}(\xb-\thetab_n)||)\right|^p d\nu(\xb) \\
     & \qquad + \lim_{n\rightarrow\infty} \int_{\R^d}\left|h(|| \Sigmab^{-1/2}(\xb-\thetab)||) - h_{\epsilon}(|| \Sigmab^{-1/2}(\xb-\thetab)||)\right|^p d\nu(\xb)\\
     &\qquad  + \lim_{n\rightarrow\infty} \int_{\R^d}\left|h_{\epsilon}(|| \Sigmab_n^{-1/2}(\xb-\thetab_n)||) - h_{\epsilon}(|| \Sigmab^{-1/2}(\xb-\thetab)||)\right|^p d\nu(\xb)\\
     & \leq 2\epsilon^p + \lim_{n\rightarrow\infty} \int_{\R^d}\left|h_{\epsilon}(|| \Sigmab_n^{-1/2}(\xb-\thetab_n)||) - h_{\epsilon}(|| \Sigmab^{-1/2}(\xb-\thetab)||)\right|^p d\nu(\xb).
   \end{align*}
   Given that $h_{\epsilon}\in C_{c}^{\infty}(\R^+)$,   Lebesgue's dominated convergence theorem implies that the latter limit is zero.  Now, for all $\epsilon>0$, 
   $$\lim_{n\rightarrow\infty} \int_{\R^d} \left|h(|| \Sigmab_n^{-1/2}(\xb-\thetab_n)||)-h(|| \Sigmab^{-1/2}(\xb-\thetab)||)\right|^pd\nu(\xb)<2\epsilon^p.$$
    The claim follows.  \hfill$\square$
   
   We now turn to the proof of Theorem~\ref{asympt}. 
   
   \subsection*{Proof of Theorem~\ref{asympt}}
 {\it (i)} To start with, let us show that  
  \begin{align*}
      (\Deltab_{3}(\hat\varthetab_0))'(\Gammab_{f;33}(\varthetab_0))^{-1}\Deltab_{3}(\hat\varthetab_0) - (\Deltab_{3}(\varthetab_0))'(\Gammab_{f;33}(\varthetab_0))^{-1}\Deltab_{3}(\varthetab_0) = o_{\rm P}(1)
  \end{align*}
as $n\rightarrow\infty$ under $ {\rm P}^{(n)}_{\thetab,\,\Sigmab,\,\zerob;f}$. The asymptotic linearity property combined with Lemma 4.4 of \citet{kreiss1987adaptive} entails that $\Deltab_{3}(\hat\varthetab_0) - \Deltab_{3}(\varthetab_0) = o_{\rm P}(1)$ as $n\rightarrow\infty$. Hence,  
$$(\Deltab_{3}(\hat\varthetab_0))'(\Gammab_{f;33}(\varthetab_0))^{-1}\Deltab_{3}(\hat\varthetab_0) = (\Deltab_{3}(\varthetab_0))'(\Gammab_{f;33}(\varthetab_0))^{-1}\Deltab_{3}(\varthetab_0) + o_{\rm P}(1)$$ as $n\rightarrow\infty$ 
 under $ {\rm P}^{(n)}_{\thetab,\,\Sigmab,\,\zerob;f}$ and the asymptotic normality of $\Deltab_{3}(\varthetab_0)$   yields the desired result for given $f$. This, however, holds for any $f\in\mathcal{F}_1$ and $\Sigmab\in\mathcal{S}_d$, so that {\it (i)} follows under the entire $\mathcal{H}_{0;\thetab}$.

 {\it (ii)} By contiguity, 
 $$(\Deltab_{3}(\hat\varthetab_0))'(\Gammab_{f;33}(\varthetab_0))^{-1}\Deltab_{3}(\hat\varthetab_0) - (\Deltab_{3}(\varthetab_0))'(\Gammab_{f;33}(\varthetab_0))^{-1}\Deltab_{3}(\varthetab_0)=o_{\rm P}(1)$$
  under $ {\rm P}^{(n)}_{\thetab,\Sigmab,n^{-1/2}\taub^{(n)}_3;g,\Pi}$ for every $\Sigmab$. By the Central Limit Theorem,  under~$ {\rm P}^{(n)}_{\thetab,\,\Sigmab,\,\zerob;g}$ and as $n\rightarrow\infty$ \small
\begin{align*}
&\left(\hspace{-35mm}
\begin{array}{c}
\Deltab_{3}(\thetab,\Sigmab,\zerob)\\
(\taub^{(n)}_2)'\Deltab_{g;2}(\thetab,\Sigmab,\zerob) +(\taub^{(n)}_3)'\Deltab_{3}(\thetab,\Sigmab,\zerob)\\ 
\qquad\qquad\qquad\qquad\qquad\qquad-\frac{1}{2}(\taub^{(n)}_2)'\Gammab_{g;22}(\varthetab_{\pmb{0}}) \taub^{(n)}_2
 -\frac{1}{2}(\taub^{(n)}_3)' \Gammab_{g;33}(\varthetab_{\pmb{0}})   \taub^{(n)}_3 +o_{\rm P}(1)
\end{array}\right) \\
&\qquad \stackrel{\mathcal{D}}{\longrightarrow}\mathcal{N}_{d+1}\left(\left(
\begin{array}{c}
\zerob\\
0
\end{array}
\right),\left( 
\begin{array}{cc}
4(\dot{\Pi}(0))^2 \Ib_d&4(\dot{\Pi}(0))^2\taub_3\\
4(\dot{\Pi}(0))^2\taub_3'&\taub_2'\Gammab_{g;22}(\thetab,\Sigmab,\zerob)\taub_2 +\taub_3'\taub_3 4(\dot{\Pi}(0))^2
\end{array}\right)\right)
\end{align*}\normalsize
for   $\taub_3=\lim_{n\rightarrow\infty}\taub^{(n)}_3$ and $\taub_2=\lim_{n\rightarrow\infty}\taub^{(n)}_2$. 
 The asymptotic distribution of$(\Gammab_{f;33}(\varthetab_{\pmb{0}}))^{-1/2}\Deltab_{3}(\thetab,\Sigmab,\zerob)$ under the alternative then follows from Le Cam's Third Lemma. 

{\it (iii)}  The asymptotic level~$\alpha$ of $\phi^{(n)}_{\thetab}$ under $\mathcal{H}_{0;\thetab}$ follows from the asymptotic normality provided under {\it (i)}. Local asymptotic maximinity is a consequence of the weak convergence to Gaussian shifts of the local skewness experiments.~\hfill$\square$

\section*{Appendix C: Proof of Theorem~4.1, Lemma~4.1, and Lemma~4.2}

\subsection*{Proof of Theorem~\ref{asympt2}}
 {\it (i)}  Let us show that 
\begin{equation}
      (\Deltab^\dagger_{f;3}(\hat\varthetab_0))'\!\left(\Gammab^\dagger_{f;33}(\hat\varthetab_0)\right)^{-1}\!\Deltab^\dagger_{f;3}(\hat\varthetab_0)   
    \!  -\! ({\Deltab}^\dagger_{f;3}(\varthetab_0))'\!\left({\Gammab}^\dagger_{f;33}(\varthetab_0)\right)^{-1}\!{\Deltab}^\dagger_{f;3}(\varthetab_0)\label{label}
  \end{equation}
  is $ o_{\rm P}(1)$ as $n\rightarrow\infty$ under $ {\rm P}^{(n)}_{\thetab,\,\Sigmab,\,\zerob;f}$. Continuity of the Fisher information matrices and asymptotic linearity yield 
\begin{eqnarray*} 
{\Deltab}^\dagger_{f;3}(\hat\thetab,\hat\Sigmab,\zerob)\!&\!\! =\!\! &\! \Deltab_{3}(\hat\thetab,\hat\Sigmab,\zerob) - \Gammab_{f;13}(\hat\varthetab_0) \Gammab_{f;11}^{-1}(\hat\varthetab_0) \Deltab_{f;1}(\hat\thetab,\hat\Sigmab,\zerob)\\
\!&\!\! =\!\! &\! \Deltab_{3}(\hat\thetab,\hat\Sigmab,\zerob) - \Gammab_{f;13}(\varthetab_0) \Gammab_{f;11}^{-1}(\varthetab_0) \Deltab_{f;1}(\hat\thetab,\hat\Sigmab,\zerob)+o_{\rm P}(1)\\
\!&\!\! =\!\! &\! \Deltab_{3}(\thetab,\Sigmab,\zerob) - \Gammab_{f;13}(\varthetab_0)n^{1/2}(\hat\thetab-\thetab)- \Gammab_{f;13}(\varthetab_0) \Gammab_{f;11}^{-1}(\varthetab_0) \Deltab_{f;1}(\thetab,\Sigmab,\zerob)\\
\!&\!\! +\!\! &\! \Gammab_{f;13}(\varthetab_0) \Gammab_{f;11}^{-1}(\varthetab_0)\Gammab_{f;11}(\varthetab_0)n^{1/2}(\hat\thetab-\thetab)+o_{\rm P}(1)\\
\!&\!\! =\!\! &\!\Deltab_{3}(\thetab,\Sigmab,\zerob)- \Gammab_{f;13}(\varthetab_0) \Gammab_{f;11}^{-1}(\varthetab_0) \Deltab_{f;1}(\thetab,\Sigmab,\zerob)+o_{\rm P}(1)\\
\!&\!\! =\!\! &\!  {\Deltab}^\dagger_{f;3}(\thetab,\Sigmab,\zerob)+o_{\rm P}(1)
  \end{eqnarray*}
as $n\rightarrow\infty$ under $ {\rm P}^{(n)}_{\thetab,\,\Sigmab,\,\zerob;f}$. The continuous mapping theorem implies that~${\Gammab}^\dagger_{f;33}(\hat\varthetab_0) -{\Gammab}^\dagger_{f;33}(\varthetab_0) = o_{\rm P}(1)$, so that~$\Big({\Gammab}^\dagger_{f;33}(\hat\varthetab_0)\Big)^{-1} -\big({\Gammab}^\dagger_{f;33}(\varthetab_0) \Big)^{-1}$ is~$o_{\rm P}(1)$ under $ {\rm P}^{(n)}_{\thetab,\,\Sigmab,\,\zerob;f}$.  A simple application of Slutsky's Lemma then yields the desired result  that~\eqref{label} is $o_{\rm P}(1)$;  the   asymptotic normality of ${\Deltab}^\dagger_{f;3}(\varthetab_0)$  completes the proof of this part of the theorem.  \smallskip
  
 {\it (ii)}  By contiguity, 
 $$ ({\Deltab}^\dagger_{f;3}(\hat\varthetab_0))'({\Gammab}^\dagger_{f;33}(\hat\varthetab_0))^{-1}{\Deltab}^\dagger_{f;3}(\hat\varthetab_0) - ({\Deltab}^\dagger_{f;3}(\varthetab_0))'({\Gammab}^\dagger_{f;33}(\varthetab_0))^{-1}{\Deltab}^\dagger_{f;3}(\varthetab_0) =o_{\rm P}(1)$$ under $ {\rm P}^{(n)}_{\thetab,\Sigmab,n^{-1/2}\taub^{(n)}_3;f,\Pi}$ for every $\thetab$ and  $\Sigmab$. The Central Limit Theorem entails   
    \begin{align*}
  \displaystyle
&\left(\begin{array}{c}
{\Deltab}^\dagger_{f;3}(\thetab,\Sigmab,\zerob)\\
(\taub^{(n)})'((\Deltab_{f;1}(\varthetab_0))', (\Deltab_{f;2}(\varthetab_0))', (\Deltab_{3}(\varthetab_0))')'-\frac{1}{2}(\taub^{(n)})'\Gammab_f(\varthetab_0)\taub^{(n)}+o_{\rm P}(1)
\end{array}\right) \\
&\qquad\quad  \stackrel{\mathcal{D}}{\rightarrow}\mathcal{N}_{d+1}\left(\left(\!
\begin{array}{c}
\zerob\\
0
\end{array}
\!\right)\! ,\left( \!\!
\begin{array}{cc}
\displaystyle
4(\dot{\Pi}(0))^2\frac{ \mathcal{I}_{d,f}- d}{\mathcal{I}_{d,f}}\Ib_d& \displaystyle 4(\dot{\Pi}(0))^2\frac{ \mathcal{I}_{d,f} - d}{\mathcal{I}_{d,f}}\taub_3\\
\displaystyle
4(\dot{\Pi}(0))^2\frac{ \mathcal{I}_{d,f} - d}{\mathcal{I}_{d,f}}\taub_3'&  \taub'\Gammab_{f}(\varthetab_0)\taub\\ 
\end{array}\right)\right)
\end{align*}
under $ {\rm P}^{(n)}_{\thetab,\,\Sigmab,\,\zerob;f}$  for $\taub=(\taub_1',\taub_2',\taub_3')'$ with $\taub_j=\lim_{n\rightarrow\infty}\taub^{(n)}_j$ for $j=1,2,3$. 
 The asymptotic distribution of $(\Gammab^\dagger_{f;33}(\varthetab_0))^{-1/2}{\Deltab}^\dagger_{f;3}(\thetab,\Sigmab,\zerob)$ under the alternative   follows from Le Cam's Third Lemma. 
  
{\it (iii)} The asymptotic level~$\alpha$ of $\phi^{(n)}_f$ under $\mathcal{H}_{0;f}$ follows from the asymptotic normality provided under (i). Local asymptotic maximinity is a consequence of the weak convergence to Gaussian shifts of the local skewness experiments.~\hfill$\square$

\subsection*{Proof of Lemma~\ref{Lemma1}}
Rewrite the difference $ \widehat{\mathcal{K}}_{d,f}(\hat\thetab,\hat\Sigmab)  -\mathcal{K}_{d,f,g}$ as 
$$ \widehat{\mathcal{K}}_{d,f}(\hat\thetab,\hat\Sigmab) - \widehat{\mathcal{K}}_{d,f}(\thetab,\Sigmab) + \widehat{\mathcal{K}}_{d,f}(\thetab,\Sigmab) -\mathcal{K}_{d,f,g}.$$
The Law of Large Numbers implies that   $ \widehat{\mathcal{K}}_{d,f}(\thetab,\Sigmab) -\mathcal{K}_{d,f,g}=o_{\rm P}(1)$ as~$n\rightarrow\infty$ under $ {\rm P}^{(n)}_{\thetab,\,\Sigmab,\,\zerob;g}$. Letting   $h(r)= \varphi_f'(r) + \frac{d-1}{r}\varphi_f(r)$ in Lemma~\ref{glemma} with~$p=1$ (integrability w.r.t.~$r^{d-1}g(r)dr$ holds since $g\in\mathcal{F}_{1;f}$), we get the~$L^1$-convergence to zero  of $ \widehat{\mathcal{K}}_{d,f}(\thetab_n,\Sigmab_n) - \widehat{\mathcal{K}}_{d,f}(\thetab,\Sigmab)$, hence also
\begin{equation}\label{convK}
 \widehat{\mathcal{K}}_{d,f}(\thetab_n,\Sigmab_n) - \widehat{\mathcal{K}}_{d,f}(\thetab,\Sigmab)=o_{\rm P}(1)\quad\mbox{as}\,\,n\rightarrow\infty\,\,\mbox{under}\,\,{\rm P}^{(n)}_{\thetab,\,\Sigmab,\,\zerob;g}.
\end{equation} This, combined with Lemma 4.4 of \citet{kreiss1987adaptive},  concludes the proof. \hfill$\square$

\subsection*{Proof of Lemma~\ref{Lemma2}}
{\it (i)} Rewrite ${\Deltab}^\ddagger_{f;3}(\hat\varthetab_0)-{\Deltab}^\ddagger_{fg;3}(\varthetab_0)$ as
 $${\Deltab}^\ddagger_{f;3}(\hat\varthetab_0) - {\Deltab}^\ddagger_{f;3}(\varthetab_0) + {\Deltab}^\ddagger_{f;3}(\varthetab_0) -{\Deltab}^\ddagger_{fg;3}(\varthetab_0).$$
We have 
 \begin{align*}{\Deltab}^\ddagger_{f;3}(\varthetab_0) -&{\Deltab}^\ddagger_{fg;3}(\varthetab_0) \\
 =&
 -2\dot{\Pi}(0)dn^{-1/2}\sum_{i=1}^n\varphi_f(d_i(\thetab,\Sigmab))\Ub_i(\thetab,\Sigmab)\left(\frac{1}{ \widehat{\mathcal{K}}_{d,f}(\thetab,\Sigmab)}-\frac{1}{\mathcal{K}_{d,f,g}}\right).
 \end{align*} 
 The continuous mapping theorem combined with the Law of Large Numbers, the fact that $\mathcal{K}_{d,f,g}\neq0$, and the integrability of $\varphi_f$ w.r.t. $r^{d-1}g(r)dr$ yield 
 $${\Deltab}^\ddagger_{f;3}(\varthetab_0) -{\Deltab}^\ddagger_{fg;3}(\varthetab_0)=o_{\rm P}(1)$$ as $n\rightarrow\infty$ under ${\rm P}^{(n)}_{\thetab,\,\Sigmab,\,\zerob;g}$. 
 
  Next, let us show that ${\Deltab}^\ddagger_{f;3}(\hat\varthetab_0) - {\Deltab}^\ddagger_{f;3}(\varthetab_0) = o_{\rm P}(1)$ as $n\rightarrow\infty$ under~${\rm P}^{(n)}_{\thetab,\,\Sigmab,\,\zerob;g}$.
Therefore,  note that (in view of the existence of finite second-order moments)
$$ \displaystyle n^{-1/2}\sum_{i=1}^n\left[\din(\thetab_n,\Sigmab_n)\Ubin(\thetab_n,\Sigmab_n) - \din(\thetab,\Sigmab)\Ubin(\thetab,\Sigmab)\right] = o_{L^2}(1)$$
as $n\rightarrow\infty$ under ${\rm P}^{(n)}_{\thetab,\,\Sigmab,\,\zerob;g}$, which directly implies the convergence in probability.  Let us show that, similarly, 
\begin{align*}& n^{-1/2}\sum_{i=1}^n\left[\frac{1}{ \widehat{\mathcal{K}}_{d,f}(\thetab_n,\Sigmab_n)}\varphi_f(\din(\thetab_n,\Sigmab_n))\Ubin(\thetab_n,\Sigmab_n)\right. \\ 
&\qquad\qquad\qquad\qquad\qquad\left.  - \frac{1}{ \widehat{\mathcal{K}}_{d,f}(\thetab,\Sigmab)}\varphi_f(\din(\thetab,\Sigmab))\Ubin(\thetab,\Sigmab) \right] = o_{\rm P}(1).
\end{align*}
The latter expression can be rewritten as
\begin{align}\nonumber
   & \left[\frac{1}{ \widehat{\mathcal{K}}_{d,f}(\thetab_n,\Sigmab_n)}- \frac{1}{ \widehat{\mathcal{K}}_{d,f}(\thetab,\Sigmab)}\right]n^{-1/2}\sum_{i=1}^n\varphi_f(\din(\thetab_n,\Sigmab_n))\Ubin(\thetab_n,\Sigmab_n)  \\
  & \displaystyle
   \qquad\quad +\frac{n^{-1/2}}{ \widehat{\mathcal{K}}_{d,f}(\thetab,\Sigmab)}  \sum_{i=1}^n\left[\varphi_f(\din(\thetab_n,\Sigmab_n))\Ubin(\thetab_n,\Sigmab_n)\! - \!\varphi_f(\din(\thetab,\Sigmab))\Ubin(\thetab,\Sigmab)\right].\label{secondt}
\end{align}
Combined with the continuous mapping theorem,~\eqref{convK}  implies that~$\displaystyle \frac{1}{ \widehat{\mathcal{K}}_{d,f}(\thetab_n,\Sigmab_n)}- \frac{1}{ \widehat{\mathcal{K}}_{d,f}(\thetab,\Sigmab)}$ is $ o_{\rm P}(1)$ as $n\rightarrow\infty$ under~${\rm P}^{(n)}_{\thetab,\,\Sigmab,\,\zerob;g}$, which takes care of the first term in \eqref{secondt}  provided that 
$$n^{-1/2}\sum_{i=1}^n\varphi_f(\din(\thetab_n,\Sigmab_n))\Ubin(\thetab_n,\Sigmab_n) = O_{\rm P}(1)$$ as $n\rightarrow\infty$ under ${\rm P}^{(n)}_{\thetab,\,\Sigmab,\,\zerob;g}$. This fact, however, follows from the Central Limit Theorem applied to $n^{-1/2}\sum_{i=1}^n\varphi_f(\din(\thetab,\Sigmab))\Ubin(\thetab,\Sigmab)$ and the $L^2$ convergence to zero of 
$$n^{-1/2} \sum_{i=1}^n\left[\varphi_f(\din(\thetab_n,\Sigmab_n))\Ubin(\thetab_n,\Sigmab_n) - \varphi_f(\din(\thetab,\Sigmab))\Ubin(\thetab,\Sigmab)\right],$$
which we shall establish now (that proof is also required for showing that the second term above is $o_{\rm P}(1)$). It is sufficient to show that 
\begin{align*}
\nonumber& {\rm E}\left[\|\varphi_f(\din(\thetab_n,\Sigmab_n))\left[\Ubin(\thetab_n,\Sigmab_n) -\Ubin(\thetab,\Sigmab)\right] \right.
\\
 &\left. \label{first summand}\quad\quad\qquad+ \left[\varphi_f(\din(\thetab_n,\Sigmab_n))-\varphi_f(\din(\thetab,\Sigmab))\right]\Ubin(\thetab,\Sigmab) \|^2\right]=:E_1 + E_2=o(1)
\end{align*}
 as $n\rightarrow\infty$ under ${\rm P}^{(n)}_{\thetab,\,\Sigmab,\,\zerob;g}$. Applying H\"older's inequality for $p= {(2+\epsilon)}/{2}$ and~$q={(2+\epsilon)}/{\epsilon}$, then using the fact that $g\in\mathcal{F}_{1;f}$ and  $\|\Ubin(\thetab_n,\Sigmab_n)\|\leq 1$, 
 together with  Lebesgue's dominated convergence theorem, one easily obtains that~$E_1=o(1)$. 
 The  convergence to zero of $E_2$ follows from Lemma~\ref{glemma} with~$h(r) = \varphi_f(r)$ and $p=2$ (integrability w.r.t.\ $r^{d-1}g(r)dr$ holds \linebreak  for~$g\in\mathcal{F}_{1;f}$). Since $L^2$ convergence implies convergence in probability, the Law of Large Numbers and the continuous mapping theorem applied to~${ \widehat{\mathcal{K}}_{d,f}^{-1}(\thetab,\Sigmab)}$ complete the proof of part {\it (i)}. 

{\it (ii)} We still have to show that $\widehat{\Gammab}^\ddagger_{f}(\hat\varthetab_0)-{\Gammab}^\ddagger_{f}(\varthetab_0) = o_{\rm P}(1)$ as $n\rightarrow\infty$ under~$ {\rm P}^{(n)}_{\thetab,\,\Sigmab,\,\zerob;g}$. In view of Lemma 4.4 of \citet{kreiss1987adaptive}, this reduces to proving that $\widehat{\Gammab}^\ddagger_{f}(\varthetab_{0n})-{\Gammab}^\ddagger_{f}(\varthetab_0) = o_{\rm P}(1).$  The latter rewrites  as 
$$\widehat{\Gammab}^\ddagger_{f}(\varthetab_{0n}) - {\Gammab}^\ddagger_{f}(\varthetab_{0n}) +{\Gammab}^\ddagger_{f}(\varthetab_{0n}) -{\Gammab}^\ddagger_{f}(\varthetab_0)=: \frac{4(\dot\Pi(0))^2}{d}({\bf A}_1 + {\bf A}_2).$$
Term ${\bf A}_1$ takes the form
\begin{align*}
{\bf A}_1=  &\displaystyle
-\frac{2d}{n}\sum_{i=1}^n\din(\thetab_n,\Sigmab_n)\varphi_f(\din(\thetab_n,\Sigmab_n)) \left(\frac{1}{ \widehat{\mathcal{K}}_{d,f}(\thetab_n,\Sigmab_n)}-\frac{1}{\mathcal{K}_{d,f,g}}\right) \\
 \displaystyle
&+\frac{d^2}{n}\sum_{i=1}^n(\varphi_f(\din(\thetab_n,\Sigmab_n)))^2\left(\frac{1}{( \widehat{\mathcal{K}}_{d,f}(\thetab_n,\Sigmab_n))^2}-\frac{1}{(\mathcal{K}_{d,f,g})^2}\right).
\end{align*}
The proof of Lemma~\ref{Lemma1},  combined with the continuous mapping theorem, implies that both 
$$ \displaystyle  \frac{1}{ \widehat{\mathcal{K}}_{d,f}(\thetab_n,\Sigmab_n)}-\frac{1}{\mathcal{K}_{d,f,g}} \ \ \text{and} \ \  \displaystyle \frac{1}{( \widehat{\mathcal{K}}_{d,f}(\thetab_n,\Sigmab_n))^2}-\frac{1}{(\mathcal{K}_{d,f,g})^2} $$
 are $o_{\rm P}(1)$ as~$n\rightarrow\infty$ under ${\rm P}^{(n)}_{\thetab,\,\Sigmab,\,\zerob;g}$.
Using similar arguments as above, one can show  that 
$$\frac{1}{n}\sum_{i=1}^n\din(\thetab_n,\Sigmab_n)\varphi_f(\din(\thetab_n,\Sigmab_n))-\frac{1}{n}\sum_{i=1}^n\din(\thetab,\Sigmab)\varphi_f(\din(\thetab,\Sigmab)) =o_{L^1}(1),$$ hence that~$\frac{1}{n}\sum_{i=1}^n\din(\thetab_n,\Sigmab_n)\varphi_f(\din(\thetab_n,\Sigmab_n))$  is $O_{\rm P}(1)$ as~$n\rightarrow\infty$ under~${\rm P}^{(n)}_{\thetab,\,\Sigmab,\,\zerob;g}$. A similar conclusion holds for~$\frac{1}{n}\sum_{i=1}^n(\varphi_f(\din(\thetab_n,\Sigmab_n)))^2$, which is also $O_{\rm P}(1)$. 
 It follows that~$\widehat{\Gammab}^\ddagger_{f}(\varthetab_{0n}) - {\Gammab}^\ddagger_{f}(\varthetab_{0n})=o_{\rm P}(1)$ as $n\rightarrow\infty$ under~$ {\rm P}^{(n)}_{\thetab,\,\Sigmab,\,\zerob;g}$.

By Lemma~\ref{glemma} with $h(r) = (r-\frac{d}{\mathcal{K}_{d,f,g}}\varphi_f(r))^2$ and $p=1$ (integrability with respect to~$r^{d-1}g(r)dr$ follows from the square integrability of~$r$ and~$\varphi_f(r)$), we get the $L^1$ convergence, hence the convergence to zero  in probability of~${\Gammab}^\ddagger_{f}(\varthetab_{0n}) -{\Gammab}^\ddagger_{f}(\varthetab_0)$  under $ {\rm P}^{(n)}_{\thetab,\,\Sigmab,\,\zerob;g}$.
\hfill$\square$

 \section*{Appendix D:  Additional numerical results} 
 
 Tables~\ref{TnotSpecdim3_1} and \ref{TnotSpecdim3_2} below are providing the finite-sample rejection frequencies, as described in Section~\ref{simuunspec}, of the unspecified-location tests: 
 our optimal tests $\phi^{\ddagger (n)}_{f}$  ($f$   elliptical Student  with $\nu=2.1$, 4,  and  8 degrees of freedom), Schott's  test~$\phi_{\text{Schott}}^{(n)}$, and Cassart's pseudo-Gaussian test~$\phi_{p{\cal G}}^{(n)}$,  in dimension   $d=3$.  
 \begin{table}[htbp]
\small
\centering
  \caption{  Rejection frequencies (out of $N=3,000$ replications), under various three-dimensional elliptical (${\lambdab = (0,0,0)}$) and related skewed densities (increasing $\lambdab$ values), of our unspecified-location optimal tests~$\phi_f^{\ddagger;(n)}$ ($f$ the trivariate elliptical $t$ distributions with~2.1,~4, and~8 degrees of freedom),  Schott's test $\phi_{\text{Schott}}^{(n)}$ and Cassart's pseudo-Gaussian test~$\phi_{p{\cal G}}^{(n)}$ for the null hypothesis of ellipticity with unspecified location. The sample size is~$n=100$, the nominal probability level~5\%.
}

{
\begin{tabular}{@{}lccccc}\toprule \toprule
  Method$\setminus \lambdab$  &     $(0,0,0)$ &$(1,-2,0)$   &    $(1,1,1)$       & $(2,2,2)$ &$(3,3,3)$     \\
\hline 
 \multicolumn{6}{c} {\footnotesize Skew-normal}   \\[.5mm]
$\phi^{\ddagger (n)}_{t_{2.1}}$ &0.044&0.110 & 0.199& 0.463& 0.558 \\ 
$\phi^{\ddagger (n)}_{t_{4}}$ &0.045& 0.118& 0.207& 0.468& 0.572 \\ 
$\phi^{\ddagger (n)}_{t_{8}}$ &0.046&0.119 & 0.211& 0.466& 0.564  \\
$\phi_{\text{Schott}}^{(n)}$ & 0.038 &0.038 &  0.038     &0.038 &  0.049 \\
$\phi_{p{\cal G}}^{(n)}$ &0.043&0.122 & 0.062& 0.083& 0.088   \\
 \multicolumn{6}{c} {\footnotesize Skew-$t_{2.1}$}   \\[.5mm]
\hline 
$\phi^{\ddagger (n)}_{t_{2.1}}$ &0.023& 0.369& 0.436& 0.571& 0.578\\
$\phi^{\ddagger (n)}_{t_{4}}$ &0.018&0.276 & 0.328& 0.446& 0.455\\
$\phi^{\ddagger (n)}_{t_{8}}$ &0.015& 0.202& 0.227 &0.320& 0.328\\
$\phi_{\text{Schott}}^{(n)}$ & 0.270 & 0.293   & 0.315  & 0.317 & 0.324  \\
$\phi_{p{\cal G}}^{(n)}$ &0.014&0.106 & 0.107& 0.158& 0.153  \\
 \multicolumn{6}{c} {\footnotesize Skew-$t_{4.1}$}   \\[.5mm]
\hline 
$\phi^{\ddagger (n)}_{t_{2.1}}$ &0.040& 0.507& 0.637& 0.845& 0.867 \\
$\phi^{\ddagger (n)}_{t_{4}}$  &0.037& 0.446& 0.557& 0.773& 0.802  \\
$\phi^{\ddagger (n)}_{t_{8}}$  &0.032&  0.364& 0.459& 0.659 &0.692  \\
$\phi_{\text{Schott}}^{(n)}$ &  0.047 & 0.042  & 0.051   & 0.055& 0.057  \\
$\phi_{p{\cal G}}^{(n)}$  &0.033& 0.224& 0.195& 0.267& 0.274	  \\
 \multicolumn{6}{c} {\footnotesize Skew-$t_{8}$}   \\[.5mm]
\hline 
$\phi^{\ddagger (n)}_{t_{2.1}}$ &0.038& 0.367& 0.507& 0.781& 0.839\\
$\phi^{\ddagger (n)}_{t_{4}}$ &0.040& 0.349 & 0.483& 0.753& 0.807\\
$\phi^{\ddagger (n)}_{t_{8}}$ &0.036& 0.316& 0.436& 0.692& 0.746\\
$\phi_{\text{Schott}}^{(n)}$ & 0.039 & 0.041& 0.034 & 0.049&    0.053 	 \\
$\phi_{p{\cal G}}^{(n)}$ &0.043&0.223 & 0.158& 0.209& 0.228  \\
 \multicolumn{6}{c} {\footnotesize Skew-$t_{10}$}   \\[.5mm]
\hline 
$\phi^{\ddagger (n)}_{t_{2.1}}$  &0.045& 0.310& 0.439& 0.724& 0.799 \\
$\phi^{\ddagger (n)}_{t_{4}}$ &0.047&0.304 & 0.420& 0.706& 0.778\\
$\phi^{\ddagger (n)}_{t_{8}}$ &0.049&0.281 &0.393& 0.658& 0.730 \\
$\phi_{\text{Schott}}^{(n)}$  & 0.034 & 0.039& 0.036 & 0.040&  0.049     \\
$\phi_{p{\cal G}}^{(n)}$ &0.054& 0.213& 0.134& 0.184& 0.208  \\
 \bottomrule \bottomrule
  \end{tabular}
  }
\label{TnotSpecdim3_1}
\end{table}
 
\begin{table}[htbp]
\small
\centering
  \caption{
Rejection frequencies (out of $N=3,000$ replications), under various three-dimensional elliptical (${\lambdab = (0,0,0)}$) and related skewed densities (increasing $\lambdab$ values), of our unspecified-location optimal tests~$\phi_f^{\ddagger;(n)}$ ($f$ the trivariate elliptical $t$ distributions with~2.1,~4, and~8 degrees of freedom),  Schott's test $\phi_{\text{Schott}}^{(n)}$ and Cassart's  pseudo-Gaussian test $\phi_{p{\cal G}}^{(n)}$ for the null hypothesis of ellipticity with unspecified location. The sample size is~$n=100$, the nominal probability level~5\%.
}
{
\begin{tabular}{@{}lccccc}\toprule \toprule
   Method$\setminus \lambdab$ &     $(0,0,0)$ &$(0.15,-2,0)$   &    $(0.15,0.15,0.15)$       & $(0.3,0.3,0.3)$ &$(0.45,0.45,0.45)$     \\
\hline 
 \multicolumn{6}{c} {\footnotesize SAS-normal}   \\[.5mm]
$\phi^{\ddagger (n)}_{t_{2.1}}$ &0.049&0.175 & 0.266& 0.840& 0.993\\
$\phi^{\ddagger (n)}_{t_{4}}$ &0.050& 0.174&0.271& 0.844& 0.991 \\ 
$\phi^{\ddagger (n)}_{t_{8}}$&0.046& 0.173& 0.263& 0.827& 0.986\\
$\phi_{\text{Schott}}^{(n)}$ &0.037 & 0.043 &  0.032     &0.032 &  0.041   \\
$\phi_{p{\cal G}}^{(n)}$ &0.054& 0.354& 0.386& 0.936 &0.998 \\
\hline   
 \multicolumn{6}{c} {\footnotesize SAS-$t_{4.1}$}   \\[.5mm]
 $\phi^{\ddagger (n)}_{t_{2.1}}$ &0.039& 0.108& 0.123& 0.400& 0.701\\
$\phi^{\ddagger (n)}_{t_{4}}$   &0.038& 0.095& 0.104&0.345& 0.606\\
$\phi^{\ddagger (n)}_{t_{8}}$  &0.035&0.083 & 0.087& 0.269& 0.489 \\
$\phi_{\text{Schott}}^{(n)}$ &  0.045 &0.040  & 0.045  & 0.049& 0.057 \\
$\phi_{p{\cal G}}^{(n)}$  &0.032& 0.124& 0.113& 0.365&  0.633\\
\hline 
 \multicolumn{6}{c} {\footnotesize LSGM}   \\[.5mm]
$\phi^{\ddagger (n)}_{t_{2.1}}$   &0.050&0.054 & 0.073& 0.182& 0.398 \\
$\phi^{\ddagger (n)}_{t_{4}}$  &0.046&0.051 & 0.069& 0.149& 0.335 \\
$\phi^{\ddagger (n)}_{t_{8}}$ &0.042& 0.047& 0.058& 0.123& 0.269   \\
$\phi_{\text{Schott}}^{(n)}$  & 0.461 &0.452& 0.450 & 0.450&  0.429   \\
$\phi_{p{\cal G}}^{(n)}$ &0.058& 0.057& 0.073& 0.132& 0.254 \\
\bottomrule  
 \multicolumn{6}{c} {\footnotesize Gaussian Mixture}   \\[.5mm]
 \multicolumn{1}{r}{$\mub_1$}  &$(0,0,0)$  &$ (0,0,0)$& $(1,0,0)$ & $(1,0,0)$& $ (1,0,0)$\\
 \multicolumn{1}{r}{$\mub_2$} &$ (0,0,0)$ & $  (-1,0,0)$& $(-1,0,0)$ & $ (-2,0,0)$ & $ (-3,0,0)$\\
 \hline 
$\phi^{\ddagger (n)}_{t_{2.1}}$  &0.044&0.436&0.905&0.955&0.957 \\
$\phi^{\ddagger (n)}_{t_{4}}$ &0.044&0.404& 0.860& 0.915&0.911  \\
$\phi^{\ddagger (n)}_{t_{8}}$ &0.044&0.365&0.777&0.832 &0.815 \\
$\phi_{\text{Schott}}^{(n)}$   & 0.081 &0.083& 0.120 & 0.236&  0.383 \\
$\phi_{p{\cal G}}^{(n)}$  &0.057&0.118&0.142&0.114&0.093  \\
 \bottomrule \bottomrule
  \end{tabular}
  }
\label{TnotSpecdim3_2}
\end{table}

 Table \ref{datatab} shows the p-values for the optimal semiparametric test $\phi^{(n)}_{\zerob}$ (specified location $\thetab_0=\zerob$) and the optimal semiparametric test for unspecified location $\phi^{\ddagger (n)}_{t_4}$, both applied to three-year subseries of the 17-dimensional  financial return data described in Section~\ref{empirsec}.

\begin{table}[ht]
\small
\centering
  \caption{
  p-values for testing for elliptical symmetry in 17-dimensional  financial return data for rolling windows over three years. We compare the optimal semiparametric test $\phi^{(n)}_{\zerob}$ for fixed $\thetab_0=\zerob$ with the optimal semiparametric test for unspecified location $\phi^{\ddagger (n)}_{t_4}$ based on the multivariate $t$ distribution with 4 degrees of freedom.
}
{
\footnotesize
\begin{tabular}{@{}ccccc}\toprule \toprule
Start&End&p-value $\phi^{\ddagger (n)}_{t_4}$ &p-value $\phi^{(n)}_{\zerob}$&Number of observations\\
 2000-01-07 &2002-12-31    &0.107659
  &0.041650
  & 778\\
 2001-01-01 &2003-12-31    &0.571561 &0.480889
 &  783\\
 2002-01-01 &2004-12-31    &0.251470
  &0.527236
  &  784\\
 2003-01-01 &2005-12-30    &0.028470
 & 0.276275
 &  783\\
 2004-01-01 &2006-12-29    &0.007923
  & 0.243174
  & 782 \\
 2005-01-03 &2007-12-31    &0.000157
 & 0.286752
 &781\\
 2006-01-02 &2008-12-31    &0.000152
  &0.100240
  &783\\
 2007-01-01 &2009-12-31    &0.000146
  & 0.183233
  &784\\
  2008-01-01 &2010-12-31   &0.005695
  & 0.872441
  & 784 \\
 2009-01-01 &2011-12-30    &0.000010
 & 0.904906
 &782 \\
 2010-01-01 &2012-12-31    & 0.000103
 & 0.626142
 & 782\\
 2011-01-03 &2013-12-31    &0.011507
 & 0.109618
 & 782\\
 2012-01-02 &2014-12-31    &0.035069
  & 0.204622
  &783\\
 2013-01-01 &2015-12-31    &0.000004
  & 0.027661
  & 783 \\
 2014-01-01 &2016-12-30    &0.000011
 & 0.380901
 &783\\
 2015-01-01 &2017-09-20    &0.006327
 & 0.766111
 & 710\\
\bottomrule \bottomrule
  \end{tabular}
  }
\label{datatab}
\end{table}

\section*{Acknowledgements}
\noindent We thank Yves Dominicy for sharing the  dataset of daily returns with us.
\newpage


\end{document}